\documentclass[apj,numberedappendix]{emulateapj}
\usepackage{natbib}
\usepackage{graphicx}
\usepackage{amsmath}
\usepackage{booktabs}
\usepackage[titletoc,toc,title]{appendix}
\usepackage{epsfig}
\usepackage{epstopdf}
\usepackage{multirow}

\newcommand{\mean}[1]{\mbox{$\langle#1\rangle$}} 
\newcommand{\msun}{\mbox{M$_\odot$}}
\newcommand\cmv{\mbox{cm$^{-3}$}}

\newcommand{\mic}{\mbox{$\mu$m}}
\newcommand{\sh}{\mbox{$\Sigma_{\text{H}_2}$}}
\newcommand{\sg}{\mbox{$\Sigma_{\text{g}}$}}
\newcommand{\ssfr}{\mbox{$\Sigma_{\text{SFR}}$}}
\newcommand{\tdep}{\mbox{$t_{\text{dep}}$}}

\begin{document}

\title{The Star Formation Relation for Regions in the Galactic Plane:
The Effect of Spatial Resolution}

\author{Nalin Vutisalchavakul \altaffilmark{1}, Neal J. Evans II \altaffilmark{1}, Cara Battersby \altaffilmark{2}}
\altaffiltext{1}{Department of Astronomy, The University of Texas at Austin, 2515 Speedway, Stop C1400, Austin, TX 78712-1205, USA }
\altaffiltext{2}{Harvard-Smithsonian Center for Astrophysics, 60 Garden St., Cambridge, MA 02138, USA}

\begin{abstract}
We examined the relations between molecular gas surface density and star formation rate surface density 
in a 11 square degree region of the Galactic Plane. 
Dust continuum at 1.1 mm from the Bolocam Galactic Plane Survey and 22 \mic\ emission from the WISE All-sky survey 
were used as tracers of molecular gas and star formation rate, respectively, across Galactic longitude of 
$31.5 \geq l \geq 20.5$ and Galactic latitude of $0.5 \geq b \geq -0.5$.
The relation was studied over a range of resolutions from  33\arcsec\ to 20\arcmin\  by convolving images to larger scales. 
The pixel-by-pixel correlation between 1.1 mm and 22 \mic\ 
increases rapidly at small scales and levels off at the scale of 5\arcmin-8\arcmin. 
We studied the star formation relation based on pixel-by-pixel analysis and 1.1 mm and 22 \mic\ peaks analysis. 
The star formation relation was found to be nearly linear with no significant changes in the form of the relation 
across all spatial scales and lie above the extragalactic relation from Kennicutt (1998). The average gas 
depletion time is $\approx 200$ Myr and does not change significantly at different scales, 
but the scatter in the depletion time decreases as the scale increases. 

\end{abstract}


\section{Introduction}

Since star formation is an important process in the formation and evolution 
of galaxies, understanding what controls the conversion of gas into stars 
is essential. 
One of the key parameters in determining the rate of star formation
(SFR) in any region should be the amount and nature of the gas
available. 
Our goal is to
use surveys of the Galactic Plane to study the relation between gas and star
formation on scales ranging from 1 to 200 pc, smaller than the scales accessible
in studies of other galaxies.

The study of the relation between gas mass and SFR goes back to the 
study of \citet{schmidt59}, who proposed a power law relation between 
SFR and the gas density. \citet{kennicutt98} studied a set of galaxies  
and found a relation between the disk-averaged SFR surface density (\ssfr)
and gas surface density (\sg) to be $\ssfr \propto \sg^{1.4}$. 
Since then, there have been many studies on the star formation relation 
in various types of galaxies, as reviewed by \citet{kennicutt12}. 

With improvements in 
instrumental resolution and sensitivity, it became possible to study the relation on 
smaller scales within individual galaxies \citep{wong02}.
Many studies used high-resolution data to study the relation down 
to sub-kiloparsec scales \citep{kennicutt07, bigiel08, 
blanc09, verley10, liu11, schruba11, leroy13}. 
Star formation has been shown to be associated with molecular gas,
rather than total gas \citep{schruba11}, and the relations between
\sh\ and \ssfr\ tend to be linear ($\ssfr \propto \sh^{1.0}$) \citep{bigiel08, 
leroy13}.
While there is a strong correlation between gas mass and SFR surface
density at disk-averaged scales, the relation showed a much larger
scatter at smaller scales, suggesting that the star
formation relation breaks down below a certain scale \citep{onodera10, 
liu11, schruba10}. 
The large scatter of the star formation relation on small scales
implies that the relation does not arise simply from propagating
the same relationship up from the scale of individual star forming regions
\citep[see][]{evans14}.
Currently, the extragalactic studies can resolve star forming regions 
down to sub-kiloparsec scales, with a few reaching scales of 
about 100 parsec \citep{schruba10, onodera10}.
 
The relations seen in the extragalactic studies are measured on 
scales that are larger than individual molecular cloud. 
To understand the physics that underly the relation, 
one should look at an individual star forming region. 
To study star formation down to scales of an individual molecular cloud, 
we need to look at closer regions.
Studies of nearby molecular clouds (distance$ < 1$ kpc) have found that star
formation was highly concentrated to dense gas
\citep{heiderman10, lada10, lada13, evans14}.
Where star formation happens, the star formation rate lies well above
the relation for extragalactic regions, indicating that the extragalactic
relation emerges from averaging over both star forming and non-star-forming
gas.  Regions of
high-mass star formation, more representative of what is probed in
studies of other galaxies, are generally more distant and require
techniques different from those used in the nearby clouds.
Studies of massive, dense regions indicated a linear relation
between SFR and dense gas \citep{wu05, wu10}, similar to
those found in other galaxies when tracers of dense gas, like HCN emission
were used \citep{gao04a, gao04b}. 
The relation using tracers of dense gas also shows larger \ssfr\ at the same \sh\
than found in extragalactic studies using all the molecular gas. 

Extragalactic studies obtain star formation relations by looking 
at regions covering large areas of the galactic disk while
star formation studies in the Milky Way have focused on individual
star forming regions that were selected based on certain criteria.
The regions studied by \citet{wu10} trace small, very dense
regions, and they were selected to have
signposts of massive star formation and thus may be biased.
To bridge the gap between the Galactic and extragalactic scales
and to understand some of the differences in the extragalactic
results, we need to study Galactic star formation on larger scales
and without the biases of previous studies.

There have been many new large scale observations of the Milky Way 
in various wavelength bands. The Spitzer legacy projects include 
the MIPSGAL \citep{carey09} and GLIMPSE \citep{churchwell09} surveys, 
giving a view of the Galactic plane in the infrared from 3.6 \mic\ to 70 \mic. 
All-sky surveys also provide information on star formation rates
in the Galactic Plane (e.g., WISE, \citealt{wright10}). 
The Bolocam Galactic Plane Survey (BGPS) observed the northern part of the 
Galactic Plane in 1.1 mm dust continuum 
(\citealt{aguirre11}, \citealt{ginsburg13} for version 2 of the data). 
The molecular gas distribution 
of the Milky Way was studied by \citet{dame01} in  $^{12}$CO
and by the Galactic Ring Survey in $^{13}$CO  \citep{jackson06},
with both studies using the $J = 1\rightarrow 0$ line. 
With these data available, we can study the relation in a larger area 
of the Milky Way and perform similar analysis on the Galactic Plane as
in the extragalactic data and study how the change in resolution and
the change in region selection method affect the result.
 Understanding the effect of the change in 
resolution and selection methods 
will be useful in comparing between Galactic and extragalactic 
studies. 

In this paper, we used the 1.1 mm dust continuum data from the 
BGPS survey as a molecular gas tracer and 22 \mic\ emission 
from WISE as a tracer of star formation to study the relation 
between gas mass and star formation for a part of the Galactic Plane. 
The details of the data used in this study are described in \S \ref{data},
and the data processing is described in \S \ref{processing}. 
We studied the relation between 
1.1 mm and 22 \mic\ emission by first looking at the pixel-by-pixel relation 
(\S \ref{pixel}) and then 
by identifying separate sources inside the regions (\S \ref{sourcebased}). 
The discussion of the results is in \S \ref{disc}, 
and the results are summarized in \S \ref{summary}.

\section{Data} \label{data}
\subsection{Emission at 1.1 mm}

Dust emission has been used as a total gas tracer in previous
studies by assuming that dust and gas are well-mixed \citep{leroy07,
 bolatto11}. The correlation
between gas and dust emission has also been studied by the 
CO survey \citep{dame01} and the recent PLANCK survey 
\citep{ade11}. \citet{ade11} argued that
dust emission is optically thin up to a column density of  $N_H \approx
10^{26}$ cm$^{-2}$ at 1 mm.
It is also less sensitive to temperature than the FIR emission.
Using dust opacities, a dust-to-gas ratio, and a typical dust temperature, 
we can estimate the gas mass.

We used the 1.1 mm dust continuum from the BGPS version 2 as a 
molecular gas tracer.
The Bolocam Galactic Plane Survey covers about 170 deg$^2$ 
in 1.1 mm continuum at an effective resolution of 33$''$ \citep{aguirre11}.
The 1.1 mm data span the range of Galactic longitude of 
$-10.5^\circ \leq l \leq 90.5^\circ$ 
and Galactic latitude of $|b| \leq 0.5^\circ$ with additional coverage in 
some selected regions \citep{ginsburg13}.
Extended sources were extracted from the 1.1 mm images to a catalog 
(Bolocat) using a watershed decomposition algorithm 
\citep{rosolowsky10, ginsburg13}. Follow-up molecular line 
observations of the sources include HCO$^+$ $J=3-2$, N$_2$H$^+$
$J=3-2$ \citep{schlingman11, shirley13}, 
and NH$_3$(1,1), (2,2), and (3,3) inversion transitions \citep{dunham11}. 
Distances to a subset of the bolocat sources are also available 
\citep{ellsworth13, ellsworth14}. 
Bolocat sources are relatively dense ($\sim 10^{3.5}$ \cmv)
structures in molecular clouds with 
angular sizes of $\approx 0.5\arcmin-2\arcmin$ \citep{ginsburg13}.

Surveys by ground-based instruments at $\lambda \sim 1$ mm
lose sensitivity to emission beyond some angular scale because it cannot
be separated from atmospheric emission variation. 
The BGPS maps completely recover emission of up to
80\arcsec\ and partially recover emission to $\approx$ 5\arcmin\ 
\citep{ginsburg13}.
These surveys thus pick
out regions of characteristic {\it volume} densities, which depend
on distance. 
\citet{dunham11} calculated sizes and other properties of a subset of 
sources and characterized the bolocat sources as cores, clumps, 
and clouds, with size scales of order 0.1 pc, 1 pc, and 10 pc,
respectively, depending on sources' distances.
As discussed later, the majority of the structures in
the regions we study here correspond to relatively dense ($n \approx
10^{3.5}$ \cmv) clumps within larger molecular clouds. 
\citet{battisti14} extracted $^{13}$CO clouds associated with
selected BGPS sources with dense gas observations and compared 
the dust mass with the mass of the parent $^{13}$CO clouds. 
Comparing a total mass of BGPS sources inside GMCs with the mass of
the GMC gave a median mass ratio of $0.11^{+0.12}_{-0.06}$. If the mass
was restricted to regions with mass surface density 
higher than 200 \msun\ pc$^{-2}$), the ratio decreased to
0.07$^{+0.13}_{-0.05}$ \citep{battisti14}. 
The dense gas mass fraction does not appear to depend on cloud mass or 
cloud mass surface density.
This result shows that the 1.1 mm sources occupy a only a small fraction of mass
and volume within the clouds. 
The maps at 1.1 mm also have a smaller chance of source confusion
along the line of sight than does CO, which traces more extended emission 
from the less dense parts of molecular clouds.

\subsection{Emission at 22 \mic} \label{iremission}
Mid-infrared continuum emission has been used as a tracer of SFR. 
The MIPSGAL survey covers our target region in 24 \mic\ MIPS band at a 
resolution of 6\arcsec. However, the saturation level is too 
low for the purpose of our large scale study. Instead, we used the 
Wide-field Infrared Survey Explorer (WISE; \citealt{wright10}) 
all-sky release images at 22 \mic\ as a tracer of star formation. 
WISE observed the entire sky in multiple exposures in four IR bands at 
3.4, 4.6, 12, and 22 \mic\ with a resolution of
12.0\arcsec\ at 22 \mic. 
SFR(24 \mic) has been calibrated using the Spitzer MIPS 24 \mic\ band. 
The WISE 22 \mic\ band overlaps the MIPS 24 \mic\ band with a slightly bluer 
response curve. The comparison between the two 
bands show that they are comparable \citep{jarrett11, anderson14}. 
We considered the difference between measurements
centered at 22 \mic\ and those centered at 24 \mic\ to be negligible,
so use relations derived for 24 \mic\ to calculate star formation rates.

The 22 \mic\ emission comes from dust heated by strong stellar 
radiation or from transiently-heated small dust grains 
\citep{draine07, calzetti07}. 
The bright emission peaks are expected to indicate
dust concentrations that are heated by high mass stars. 
The 22 \mic\ emission can then be
used to trace the presence of high mass stars and so traces the
current star formation activities. 
We used the 22 \mic\ maps to study the relation between star formation 
activities and the gas distribution. 

To convert the 22 \mic\ emission to SFR however, a conversion factor
or a calibration is needed. Several studies calibrated a relation
between MIR and SFR using extragalactic data (\citealt{calzetti07},
\citealt{calzetti10} and references therein). These calibrations were done on 
extragalactic scales assuming a fully-sampled IMF and a long 
timescale of constant star formation. 
These assumptions are not always valid when applied to smaller scales
or to regions with low mass or low SFR 
\citep{vutisalchavakul13, kennicutt12}. 
The effect of stochastically 
sampling of the IMF and the star formation history has been
studied quantitatively by \citet{dasilva12} and \citet{dasilva14}. 
They did not study the SFR measured from 24 \mic\ emission
explicitly, but they did study the SFR measured from the bolometric luminosity,
which is the closest in sensitivities to the mid-infrared emission.
These two tracers correlate closely \citep{vutisalchavakul13}.
The SFR determined from the bolometric luminosity, assuming continuous 
star formation and an IMF, has a scatter of 0.4 to 0.6 dex
for SFR$< 10^{-4}$ \msun\ yr$^{-1}$, and begins to systematically 
underestimate the actual SFR below $10^{-5}$ \msun\ yr$^{-1}$ \citep{dasilva14}. 
\citet{vutisalchavakul13} found that total infrared luminosity and 24 \mic\ 
emission can underestimate SFR by more than an order of magnitude
for molecular clouds with SFR$< 10^{-6}$ \msun\ yr$^{-1}$.
The effects of stochasticity get smaller as the total SFR of the region
increases. 
For this work, 22 \mic\ can underestimate the SFR on small scales and
especially at low surface density but  should trace 
SFR better when we look at larger scales, where we average over
multiple star forming regions and toward bright regions with high surface
density. 
  
\subsection{The Regions Studied}

The part of the Galactic plane covered in this study includes a Galactic 
longitude range of $ 20.5 \leq l \leq 31.5$ and Galactic latitude 
range of $|b| \leq 0.5$, a total of 11 deg$^2$, divided into two
equal size regions. 
Both the 1.1 mm images from BGPS version 2 and WISE 22 \mic\ images were 
combined into two separate mosaics using Montage \citep{jacob10} 
for the purpose of data
analysis: region 1 at $20.5 \leq l \leq 26.0$ and 
region 2 at $26.0 \leq l \leq 31.5$. 
Figure~\ref{fig:im} shows 1.1 mm and 22 \mic\ images for region 1 in the 
top two panels and region 2 in the third and forth panels. 
All the analysis described in the next section was performed similarly
on images for the two regions. The results were then combined for
further discussion.  
The two regions show sources with strong emission at 22 \mic,
including well-studied high mass star forming regions 
W41 and W42 in region 1 and W43 for region 2 \citep{benjamin05}. 
The end of the Galactic bar should be near the end of region 2 at 
around l $\approx 28-31^\circ$ \citep{benjamin05}.   

\section{Data Processing}\label{processing}

\subsection{Diffuse IR Background Estimation}\label{bg}
The 22 \mic\ images show diffuse background emission that 
is not necessarily associated with recent star formation. 
The diffuse (cirrus) infrared emission within our Galaxy
has long been observed and studied
\citep{low84, mivilledesch07, compiegne10}.
We constructed  diffuse emission maps to 
facilitate removal of the diffuse background in the 22 \mic\ images
for photometry and for better comparison with the 1.1 mm images where
large scale emission was automatically removed. 

The WISE 22 \mic\ images were convolved to 
a resolution of 33\arcsec\ and aligned to the BGPS images, 
which were then used for estimating the diffuse emission maps. 
To capture the variations in the background, we adopted the 
following method.
First, source areas were determined by selecting a 22 \mic\ flux contour level 
that matched bright 22 \mic\ emission areas when inspected by eye. 
All pixels inside source areas were masked as source pixels. 
Second, the images were divided into smaller rectangular subgrids at a
size of 200 by 200 pixels ($\approx 10.8\arcmin$). 
Iterative applications of Chauvenet's rejection criterion 
were performed on each subgrid by
iteratively applying a 3$\sigma$ cut until 
all the remaining pixels are within 3$\sigma$ of the average pixel value.
The average of the remaining pixel values was then taken as the
background value for the subgrid. 
Finally, subgrids with fractions of the source area above a certain clipping threshold 
were omitted. The rest of the subgrid's background values were interpolated 
with a thin plate spline interpolation to create a 
final background image. 

This method gave a reasonable representation of the diffuse emission as seen in the 
original 22 \mic\ images. We chose a source contour level, a grid size, and a clipping threshold 
that resulted in the closest approximation of the diffuse emission when inspected by eye. 
Our method gives a diffuse background image that is similar to the method of \citet{battersby11}.
More detailed descriptions on the parameters chosen and the associated 
uncertainties are provided in Appendix B, while the comparison between
two background subtraction 
methods is provided in Appendix C. 
The background-subtracted 22 \mic\ images were used for the image
convolution. 

\begin{figure}[h]
\center
\includegraphics[scale=0.8]{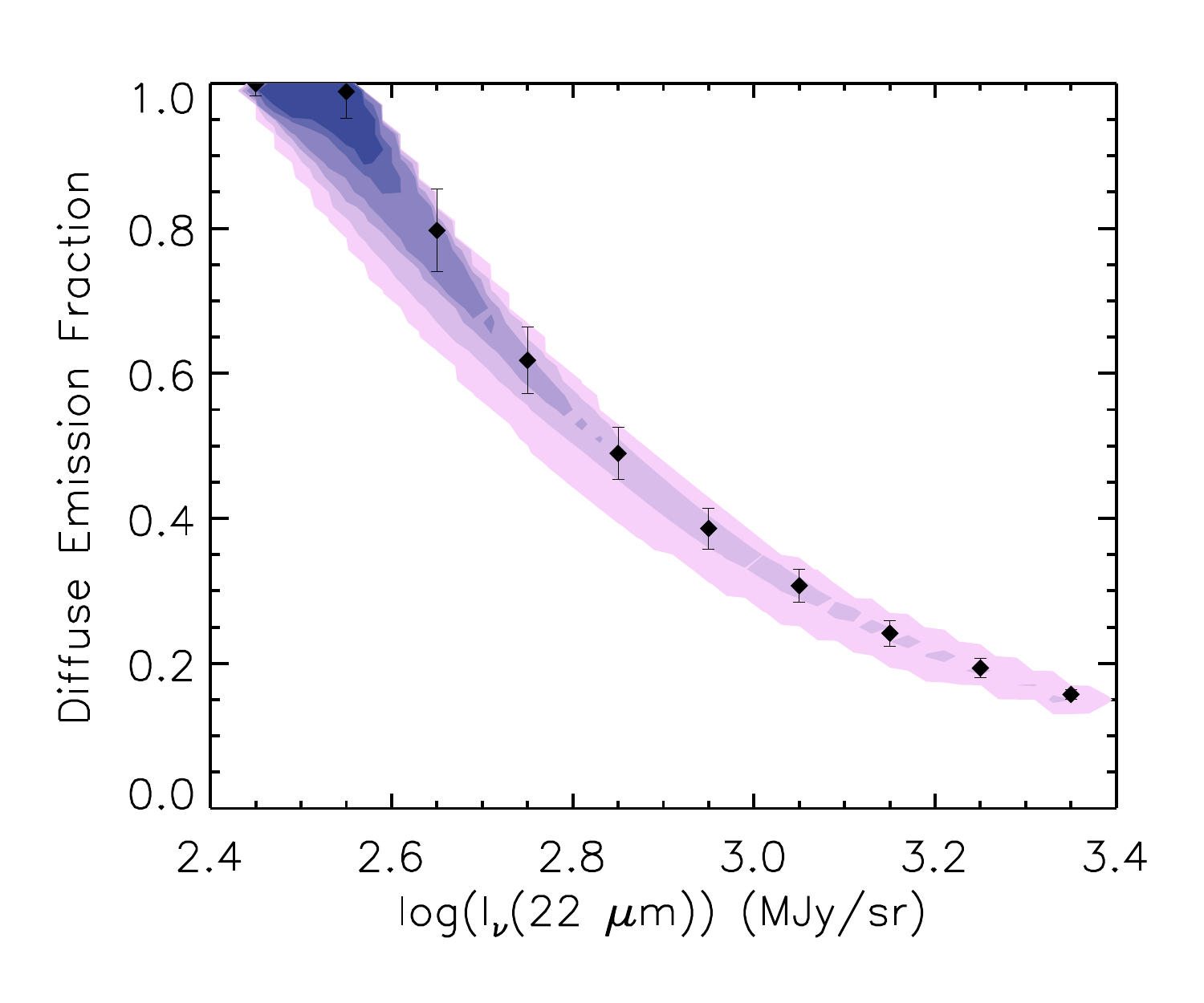}
\caption{The fraction of the diffuse (cirrus) emission in WISE 22 \mic\ image 
shows a large contribution from the diffuse emission to the total flux at low surface brightness regions.
Each data point and the error bar represent a median and a standard deviation 
of the diffuse emission over total emission inside a bin of 
log($I_\nu$(22 \mic)) with a bin size of 0.1.
The color contours represent pixel number density contours. }
\label{fig:cirrus}
\end{figure}

The result of removing diffuse emission is shown in
Figure~\ref{fig:cirrus} by comparing the fraction of the estimated
diffuse emission to total emission as a function of surface
brightness. The diffuse emission dominates the 22 \mic\ flux at 
low surface brightness and contributes a considerable fraction 
up to a surface brightness of $\approx 700$ MJy sr$^{-1}$ where 
the diffuse emission accounts for 50\% of the total emission, 
showing that removing diffuse
emission is crucial.

\subsection{Image Convolution}\label{conv}
\begin{figure}[h]
\center
\includegraphics[scale=0.9]{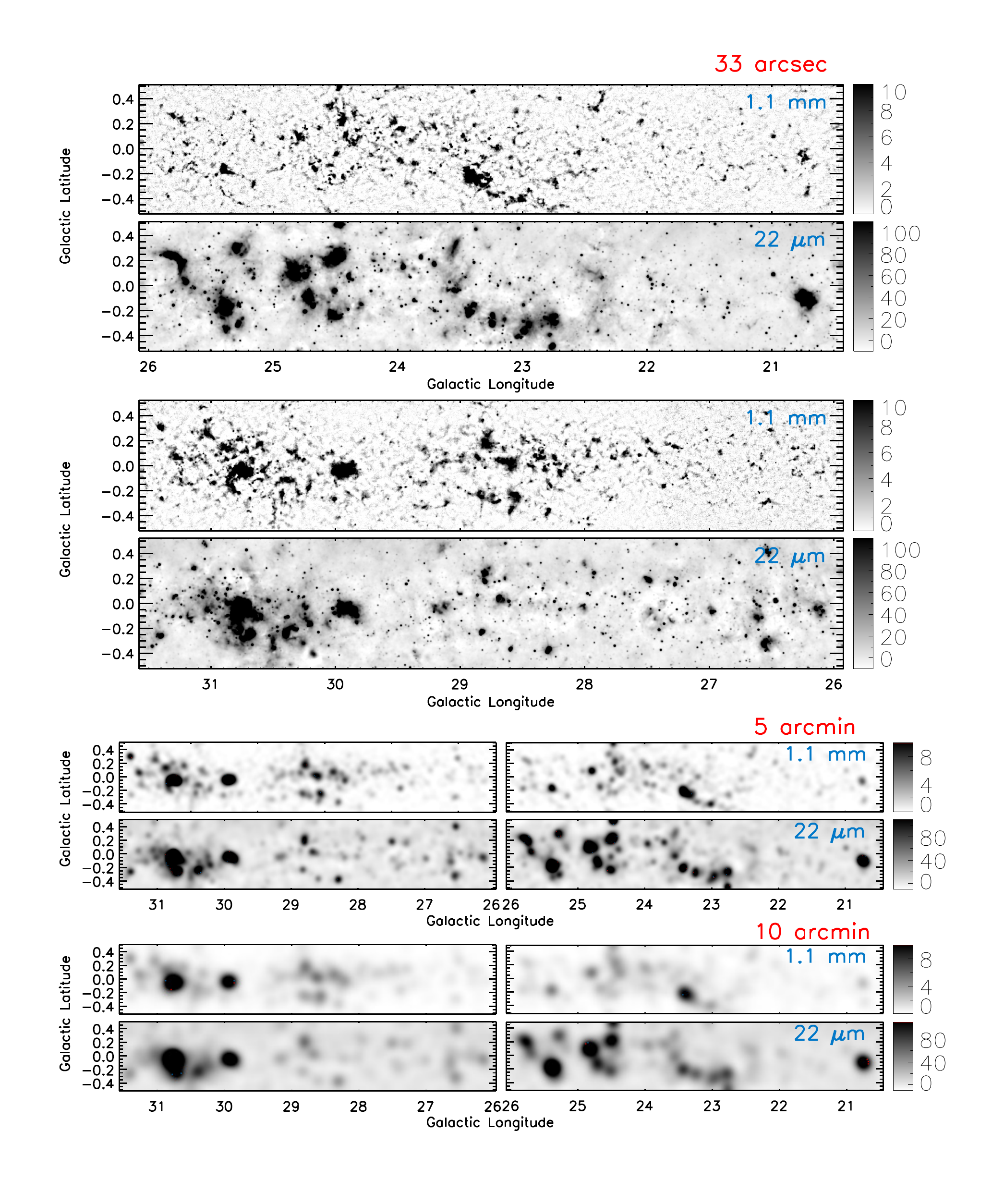}
\caption{BGPS 1.1 mm and WISE 22 \mic\ images 
for the entire region at three different resolutions of 33\arcsec\ (top 4 panels), 
5\arcmin\ (middle 4 panels), and 10\arcmin\ (bottom 4 panels).}
\label{fig:im}
\end{figure}

The original resolutions of the 1.1 mm and 22 \mic\ images were
33\arcsec\ and 12\arcsec\ respectively. 
Since we intended to study the relations between 
1.1 mm and 22 \mic\ at different scales, we 
first created a set of images at different spatial resolutions.
This was done by convolving the images with a 2D Gaussian profile of varying FWHM. 
Both the 22 \mic\ and 1.1 mm images at 
33\arcsec\ resolution were convolved with a 2D Gaussian 
kernel to resolutions of $\approx$ 1\arcmin, 2\arcmin, 3\arcmin,
 4\arcmin, 5\arcmin, 8\arcmin, 10\arcmin, 15\arcmin, and 20\arcmin. 
The scale of 20\arcmin\ was the largest scale we could achieve due to the 
limited coverage of Galactic latitude in the BGPS data. 
After the convolutions, the convolved images 
were binned to oversampling rates of $\approx$ 10 pixels per beam's FWHM.  
Figure~\ref{fig:im} shows the 1.1 mm and 22 \mic\ images at three resolutions: 33\arcsec, 
10\arcmin, and 15\arcmin. 
The whole set of the convolved images was then used for further
analysis. 

\subsection{Detection Limit}
We identified regions with unreliable detections by
 estimating the sensitivity levels of both 1.1 mm and 22 \mic\
maps. The sensitivity of the WISE 22
\mic\ images is very low compared to the average flux, and the
uncertainty in source emission is dominated by the small scale 
variations in the diffuse background emission. We estimated the 
noise in the diffuse emission based on several sky regions 
in the image at 1\arcmin\ resolution. 
The average of the standard deviation of all the sky regions 
was taken as the 1-sigma value of the
noise per pixel. The detection limits at other resolutions were estimated
by assuming that the noise drops as $1/\sqrt{N_{\text{pixels}}}$, 
where $N_{\text{pixels}}$ is the number of pixels at 1\arcmin\ scale 
inside a resolution element. Tests on the convolved images supported this
assumption.

The RMS noise for the BGPS 1.1 mm maps was estimated 
in \citet{ginsburg13} over the range of the observed Galactic
longitude. The average RMS noise in our targeted region
($20.5^\circ<l<30.5^\circ$) is $\approx 0.0021$ Jy per pixel at the 
original pixel size of 7.2\arcsec. Detection limits at other resolutions
were estimated under the same assumption as for 22 \mic\ images.

\section{Results: Pixel-by-Pixel Analysis}\label{pixel}

\subsection{Correlations between 1.1 mm and 22 \mic\ emission}
With the complete set of both 1.1 mm and 22 \mic\ images at various
scales, we looked at the correlation between the two. 
Rank correlation coefficients were calculated between 
the 1.1 mm and the 22 \mic\ flux per pixel. Rank correlation was 
used to look at the relation of the data since it is less sensitive 
to outliers and does not assume a linear relation. The result of the 
rank correlation coefficients versus resolutions is shown in
Figure~\ref{fig:pixcor}. The blue dots represent data for region 1
while the red dots represent data for region 2. 
The correlation coefficients for both regions increase rapidly
as the scale increases until the scale of $\approx$ 5\arcmin-8\arcmin,
above which they appear to be asymptoting to about 0.65 to 0.75, depending
on the region.
While some differences between the two halves of the data exist, they do not
seem large, so we combine the results from both regions in the rest of
the paper.

\subsection{The Star Formation Relation}
With the 1.1 mm and 22 \mic\ flux per pixel, we next converted 
them to molecular gas surface density (\sh) and SFR surface density
(\ssfr). 
Since the surface densities are distance independent, we can use the 
entire regions for sampling data points. 
The sampling was done by binning the images so that each pixel 
was approximately equal to a resolution element. Each pixel was 
then treated as a single data point. This method resulted in 
non-overlapping regions equal to the resolution size, 
covering the entire image, each 
with a corresponding 1.1 mm and 22 \mic\ flux surface density. 

The mass surface density can be calculated from
\begin{equation}
\Sigma_{\text{H}_2} = \left( \frac{S_\nu(1.1)}{\Omega B_\nu (T_{\text{dust}}) 
\kappa_{\text{dust},1.1}} \right) \left( \frac{\rho_g}{\rho_d} \right) , 
\end{equation}
where  $S_\nu$(1.1 mm) is the 1.1 mm flux density, $\Omega$ is the solid
angle of a pixel, $\kappa_{\text{dust},1.1}$ = 1.14 cm$^2$ g$^{-1}$
is the dust opacity at 1.1 mm per dust mass (Ossenkopf \& Henning 1994), 
and $\rho_g/\rho_d$ is the gas-to-dust mass ratio, taken to be
100 \citep{hildebrand83}.
Assuming standard values and a dust
temperature of 20 K for all sources yields
\begin{equation} \label{eq:sh}
\sh = \frac{37.2 \times S_{1.1 mm}}{\theta_{arcsec}^2} (\text{g cm}^{-2}), 
\end{equation}
where $S_\nu(1.1 mm)$ is the 1.1 mm flux density in Jansky, 
and $\theta_{arcsec}$ is the size of the region in arcsecond 
\citep{schlingman11}. 

The SFR surface density was calculated from the extragalactic
relation \citep{calzetti07}:
\begin{align} \label{eq:ssfr}
\ssfr (M_\odot \text{yr}^{-1} \text{kpc}^{-2}) = &1.56 \times 10^{-35} \nonumber \\
&[S_{24} (\text{ergs s}^{-1}~ \text{kpc}^{-2})]^{0.8104} . 
\end{align}
The 24 \mic\ luminosity surface density is described by  
\begin{equation}
S_{24} (\text{ergs s}^{-1}~ \text{kpc}^{-2})= \frac{\nu (\text{Hz}) L_{24}(\nu) 
(\text{erg s}^{-1} \text{Hz}^{-1})}{A (\text{kpc}^2)}, \nonumber
\end{equation}
where $L_{24}(\nu)$ is the 24 \mic\ luminosity per unit frequency, 
$\nu$ is the frequency, and A is the projected physical area of the region. 
 We substituted the 22 \mic\ flux for the 
24 \mic\ flux for the calculations. 
 
\begin{figure}[h]
\center
\includegraphics[scale=1]{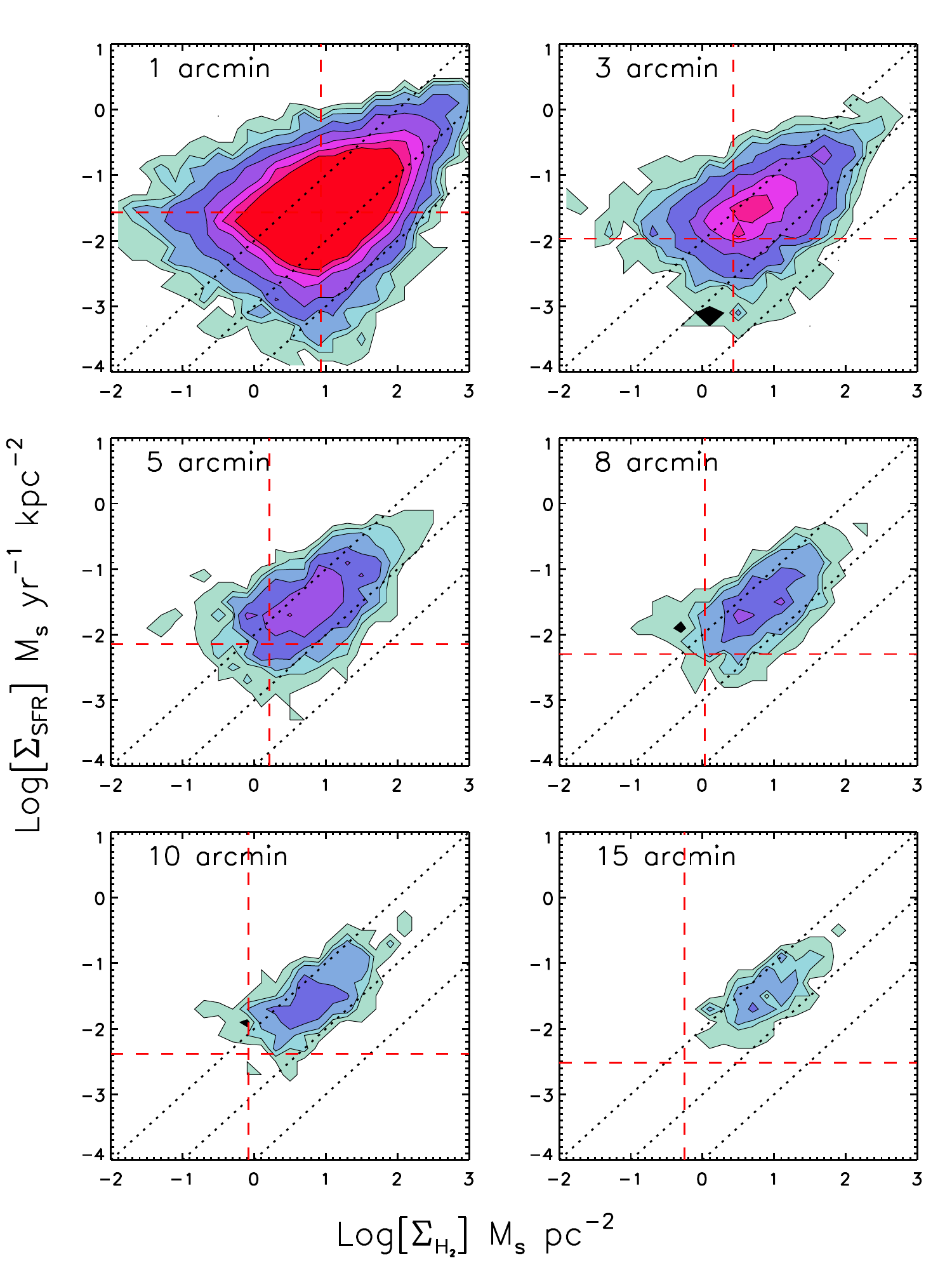}
\caption{The pixel-by-pixel star formation relations of 
\sh\ and \ssfr\ at 
different resolutions show large scatters at small 
scales and tighter realtions at larger scales. 
The contours represent 
source number densities of 
1, 3, 5, 10, 20, 40, 60, and 80 data points at the binning 
of 0.2 in \sh\ and \ssfr. 
The same color represents the same number density in all the
plots. The three dotted lines show lines of constant depletion time
with \tdep\ $= 10^{8}, 10^9, 10^{10}$ yr from top to bottom. 
The horizontal and vertical dashed lines show the 3-sigma 
detection limit for \ssfr\ and \sh\ respectively.}
\label{fig:pixsfl}
\end{figure}

The resulting \sh\ and \ssfr\ for various spatial scales are shown in 
Figure~\ref{fig:pixsfl}. 
The figure shows the star formation relations as contour plots 
at each resolution. The contours represent number densities of data
points of 1, 3, 5, 10, 20, 40, 60, and 80 data points at the binning 
of 0.2 in $\log(\sh)$ and $\log(\ssfr)$. 
The same color represents the same number density in all the plots.
The plot includes all data points with positive fluxes with 
the vertical and the horizontal dashed red lines representing detection 
(3-sigma) limits for \sh\ and \ssfr\ respectively.
Naturally, the number of data points drops as the scale gets larger,
and there are fewer pixels with values $<$ 3-sigma.

The molecular depletion time is defined to be 
\begin{equation}
\tdep = \frac{\sh}{\ssfr} \nonumber .
\end{equation}
\tdep\ can be thought of as the timescale for all the molecular gas to 
be converted into stars at the current rate of star formation. 
The three dotted lines in Figure~\ref{fig:pixsfl}
show lines of constant depletion time
with \tdep\ $= 10^{8}, 10^9, 10^{10}$ yr from top to bottom. 
The distribution of \tdep\ provides a good measure of
the scatter in the star formation relations. 

\begin{figure}[h]
\center
\includegraphics[scale=1]{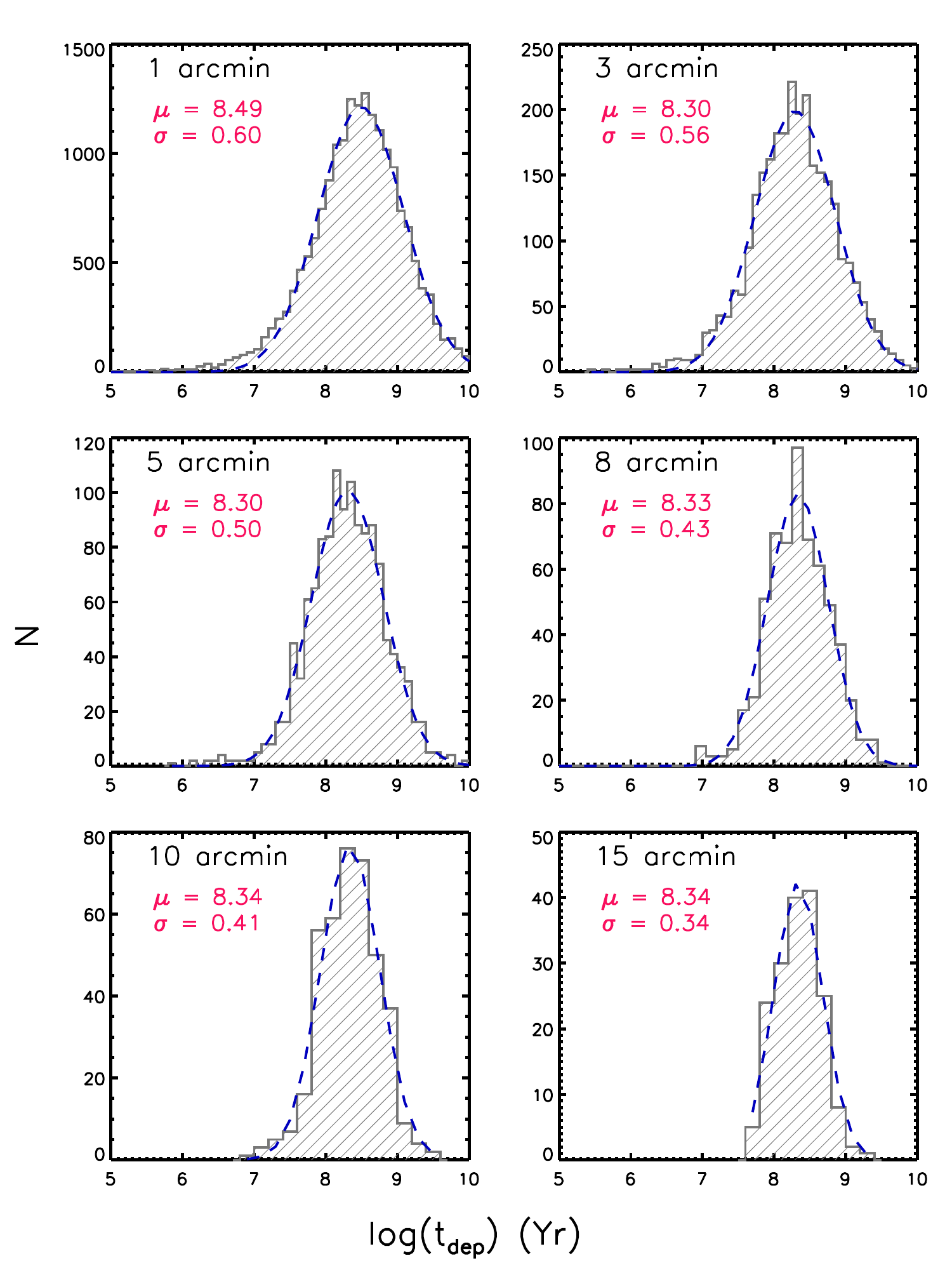}
\caption{The distributions of log(\tdep) for all the regions with
  positive 1.1 mm and 22 \mic\ flux
at the scales of 1\arcmin, 3\arcmin, 5\arcmin,
 8\arcmin, 10\arcmin, and 15\arcmin\ are shown in 
 grey.
The dashed, blue line in each plot represents
a Gaussian fit to the distribution. The mean ($\mu$) and the 
standard deviation ($\sigma$) from the Gaussian fit are shown in red for each
plot.
The mean \tdep\ from the fit does not change much at 
different scales, while the standard deviation decreases as the
scale increases. 
 }
\label{fig:pixtdep}
\end{figure}

Figure~\ref{fig:pixtdep} shows the distributions of the log(\tdep) at
various spatial scales, which include all the data points with 
positive fluxes. The dashed, blue line in each plot represents 
a Gaussian fit to the distribution. The result shows that the average 
log(\tdep) is about $8.3$, corresponding to 200 Myr, independent of
smoothing length, while the scatter, $\sigma \log(\tdep))$, 
decreases as the smoothing length increases. 

The distributions in Figure~\ref{fig:pixtdep} include many points
below the detection limit. To account for the sensitivity issue, 
we assigned the 3-sigma detection values as upper limits 
of \sh\ and \ssfr\ for all the data points with values below the detection
limits. Data points with upper limits
on both \sh\ and \ssfr\ were omitted from further analysis since 
most of them should be regions not involved in current star formation. 
Table~\ref{tdep} shows $\mean{\log(\tdep)}$ and $\sigma(\log(\tdep))$
estimated by three different methods. The first method (fit) is the
Gaussian fit to the distributions of all the data points with 
positive fluxes before assigning
upper limits as shown in Figure~\ref{fig:pixtdep}. 
The second method (limit) used the data including upper limit values 
of \sh\ and \ssfr\ to 
calculate the median and standard deviation of $\log(\tdep)$. 
The third method (EM) used an expectation-maximization algorithm 
as described in \citet{wolynetz79} to estimate the mean and standard 
deviation of a censored normal distribution. 
The EM method is based on calculating the maximum likelihood estimates 
to a distribution assumed to be normally distributed. 
The data includes left-censored (upper limit in \sh) and
right-censored (upper limit in \ssfr) data points. The uncertainties in
the EM method increase for data sets where large fractions of the
data points are censored. 
The three methods give a comparable $\mean{\log(\tdep)}$ of about 8.3
regardless of the resolution scale (Table~\ref{tdep}).
They differ more regarding $\sigma(\log(\tdep))$, but these differences
disappear at larger smoothing scales.

\section{Results: Source-based Relations}\label{sourcebased}

The pixel-by-pixel analysis provides a way to study the relation 
between \sh\ and \ssfr\ for an entire region without a need to
extract sources from the images. This method was used in some 
extragalactic studies \citep{bigiel08, liu11, leroy13}. 
However, the pixel-by-pixel analysis includes regions with low 22 \mic\ and 
1.1 mm surface densities that do not necessarily represent star formation. 
In addition, no information on distance is available.

Another approach to study the relation is to choose regions with
strong emission, as was done by some extragalactic studies 
(e.g., \citealt{kennicutt07}). 
We started by looking at extracted sources from the BGPS source
catalog. These sources also have the advantage of having additional 
properties determined for a subset of them in previous studies such as
distances, sizes, temperature, and densities. 

We started with the sources in the BGPS catalog, describing
the properties of sources in this sample (\S \ref{bgpsprop}) and study the 
star formation relation (\S \ref{bgpscat}). 
Then we considered what happens when we smoothed the images to larger scales,
and extracted data centered on 1.1 mm peaks (\S \ref{mmpeaks}). 
Finally, we compared these results to those obtained when we extracted
data centered on 22 \mic\ peaks (\S \ref{irpeaks}).

\subsection{The BGPS Source Properties}\label{bgpsprop}

The BGPS source catalog (Bolocat) provides sources extracted from the 1.1 mm 
images
along with integrated source flux and flux inside aperture diameters of 
40\arcsec, 80\arcsec, and 120\arcsec. 
These sources are possible sites of star formation, and they correspond
to cores, clumps, or clouds depending on the distance \citep{dunham11}.
Since 1.1 mm emission provides an unbiased tracer for dense molecular gas
without a pre-selected criteria, Bolocat gives a good set of sources 
for studying properties of potential star forming regions. 

Distances are not available to all the Bolocat sources, 
but kinematic distances to a subset of sources have been 
determined by \citet{ellsworth13, ellsworth14} 
using a Bayesian distance probability distribution function 
to resolve the kinematic distance ambiguity in the inner Galaxy.
The distance catalog provides distances along with the uncertainties.
In our targeted regions, 33\% of the sources have distances.
Figure \ref{fig:dis} shows the distance distribution for sources
inside our targeted region, where the distances plotted are the
maximum likelihood distances from \citet{ellsworth14}.
The shaded grey area represents a distance range inside which 90\% of 
the total number of sources reside (see \S \ref{physscales}).
The subset of the sources with distances (referred to hereafter as the
distance catalog) are generally representative
of the entire Bolocat source catalog. Their distribution in Galactic
latitude  is comparable to that of sources in the full catalog. 
However, the distance catalog is slightly biased toward sources with
larger surface brightness with a median of 6.8 and 8.0 MJy sr$^{-1}$ for
full catalog and distance catalog respectively (see the full
discussion in \citealt{ellsworth14}).

A fraction of 0.73 of the sources with distances are located in
the 5 kpc peak (between 0-7.5 kpc), and a fraction of 0.25 reside near
the 10 kpc peaks (between 8-14 kpc) while a very small fraction of sources are 
at distances around 16 kpc.  
The mean of the distance is 5.9 kpc, the median is 5.1 kpc, and the
standard deviation is 2.8 kpc.  We will use the median value to calculate
characteristic properties.
At 5 kpc, 1 arcminute corresponds to 1.45 pc, and BGPS sources
correspond to clumps  with typical densities of 10$^{3.5}$ \cmv\
\citep{dunham10, dunham11}. The sources at 10-12 kpc are better characterized
as clouds, with mean densities below 10$^{3}$ \cmv.

\subsection{Analysis Based on Bolocat Sources}\label{bgpscat}

We started the investigation at the scale of individual Bolocat sources 
by looking at the 22 \mic\ emission from these sources.
We performed aperture photometry on the 22 \mic\ images 
of 33\arcsec\ resolution at each of the Bolocat source positions 
with an aperture radius of 40\arcsec\ to compare to the 1.1 mm Bolocat flux 
of the same aperture size. 

The photometry resulted in a total of 1981 Bolocat sources in the 
whole 11 deg$^2$ region; 738 of them have a 22 \mic\ background-subtracted 
flux below the detection limit, a fraction of 0.37. 
The photometric uncertainties were determined by combining the 
observational and Poisson error from the WISE uncertainty maps 
with the estimated random uncertainties in our method of background 
subtraction, the details of which can be found in Appendix B.

We next calculated the SFR 
surface density from the 22 \mic\ flux (Equation~\ref{eq:ssfr}) and molecular gas surface 
density from the 1.1 mm flux (Equation~\ref{eq:sh}) within an  80\arcsec\ aperture. 
The typical uncertainties are $\approx$ 28\% and 7\% for
\sh\ and \ssfr\ respectively. The uncertainties for \sh\ were
determined from the calibration and photometric uncertainties
\citep{ginsburg13}.

\begin{figure}[h]
\center
\includegraphics[scale=0.8]{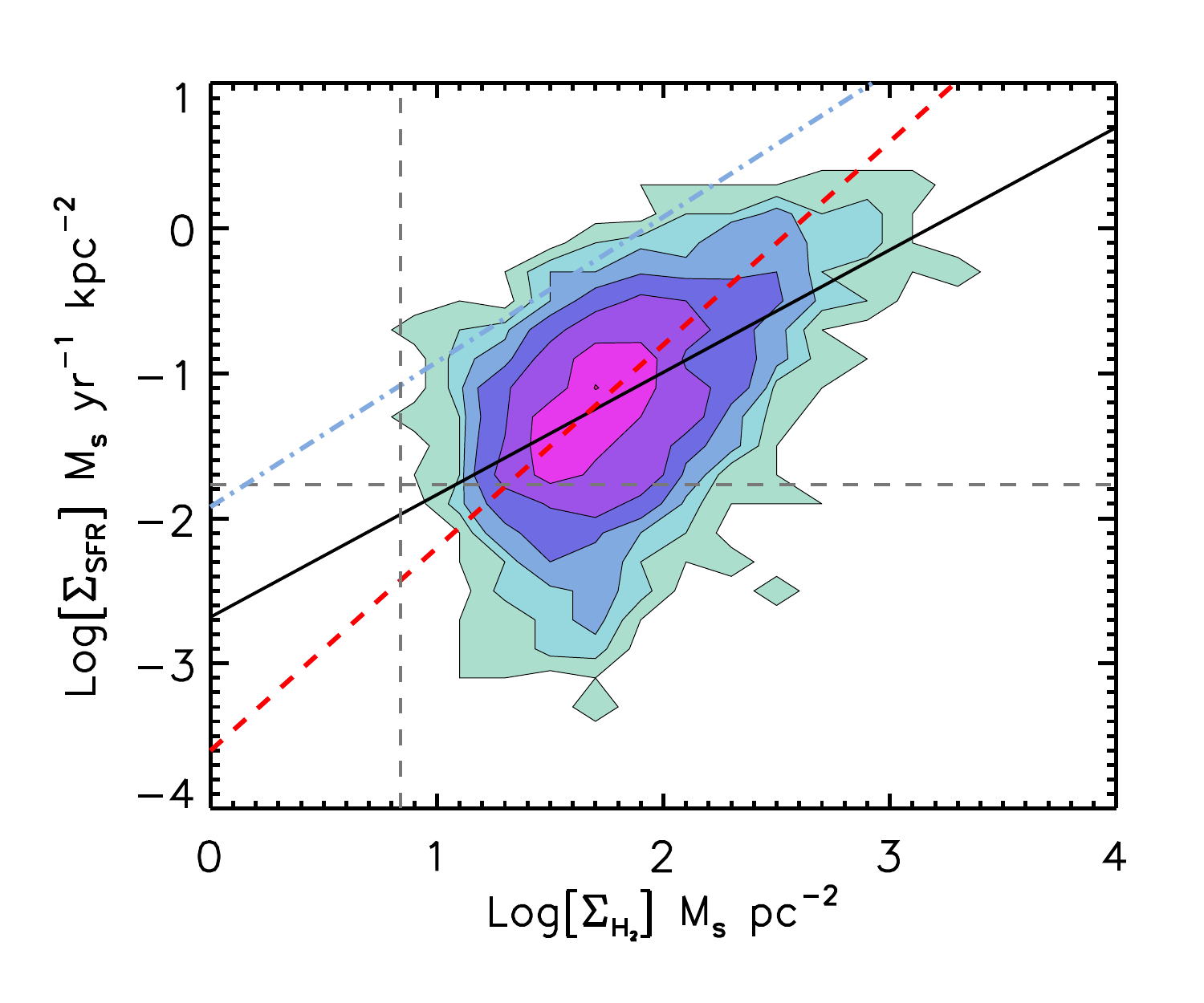}
\caption{The star formation relation for Bolocat sources with a photometry aperture 
radius of 40\arcsec shows a weak relation with a large scatter.
 The contours represent the same source number densities
as in Figure~\ref{fig:pixsfl}. The solid black 
line represents the best linear fit to the log of the data 
while the dashed, red line represents the extragalactic star formation relation 
(Equation \ref{eq:ks}), and the dot-dashed blue line represents the 
Galactic relation for dense gas from Wu et al. (2005). 
The vertical and the horizontal grey, dashed lines show the 3-sigma detection limit in \sh\ 
and \ssfr\ respectively. 
 }
\label{fig:bolocat_all}
\end{figure}

The result is shown in Figure \ref{fig:bolocat_all}. 
The contours show the source number density, 
as in Figure~\ref{fig:pixsfl}.
The data have a large scatter, with 
a rank correlation coefficient between $\log(\sh)$ and
$\log(\ssfr)$ of 0.40. 
We assumed a power law relation of the form
\begin{equation}
\ssfr \propto \sh^n ,
\end{equation}
or equivalently 
\begin{equation}
\log(\ssfr) = n \log(\sh) + a.
\end{equation} 
A linear curve fit to the log data including the uncertainties in both axes 
using MPFITEXY \citep{markwardt09}
gave fitting parameters of $n=0.84 \pm 0.03$ and $a=-2.68 \pm 0.06$,
as shown by the solid black line.
 The dashed red line represents the extragalactic star formation relation 
from \citet{kennicutt98} of 
\begin{equation} \label{eq:ks}
\ssfr = (2.5 \pm 0.7) \times 10^{-4} \sh^{1.4 \pm 0.15}.
\end{equation}
The vertical and the horizontal dashed lines 
represent the 3-sigma detection limit, 
 and the 
dot-dashed blue line represents the relation observed for dense gas as
traced by HCN from \citet{wu05}: 
\begin{equation} \label{eq:wu}
\text(SFR) (\msun\ \text{yr}^{-1}) \approx 1.2 \times 10^{-8} M_{\text{dense}}
(\msun). 
\end{equation}
A gaussian fit to the distribution of $\log(\tdep)$ gives $\log(\tdep) = 
9.0 \pm 0.48$ ($\approx 1$ Gyr).  
Including upper limits to \ssfr, the median log(\tdep) $= 9.23 \pm
0.69$. The EM method gives the mean of log(\tdep) $= 9.45 \pm 0.90$.

\subsection{Analysis based on 1.1 mm Peaks}\label{mmpeaks}
On the scale of BGPS sources, the 22 \mic\ emission
shows a weak correlation with the 1.1 mm flux with a large scatter.  
From visual inspections, the 22 \mic\ emission in the whole region is 
more diffuse than the 1.1 mm emission. A lot of the 22 \mic\ extended 
emission also does not coincide with the Bolocat source contours. 
In this section, we studied how the correlation between SFR tracers 
and gas tracers changes when we look at the regions on larger scales. 

Using the images convolved to larger angular scales (\S \ref{conv}),
we identified the local peaks of the 1.1 mm emission.
The local peaks were identified by locating pixels whose values are
larger than all the adjacent pixels. Overlapping regions were eliminated by dropping 
peaks with distances to the nearest peak less than the radius of the aperture. 
We then performed aperture photometry on 22 \mic\ and 1.1 mm images
with an aperture centered at the local 1.1 mm peaks and
 radius equal to the beam's FWHM of the 
convolved images (aperture radius = 10\arcmin\ for the images convolved to FWHM 
of 10\arcmin\ and similarly for others).  The aperture size was chosen to contain 
most of the source emission without applying aperture corrections.
The same procedures were performed on the images convolved to resolutions of 
10\arcmin, 15\arcmin, and 20\arcmin. 
Once we go to higher FWHM, the angular source sizes increase, which 
corresponds to looking at larger physical areas. At a distance of 5
kpc, 20\arcmin\ corresponds to a physical size of about 29 pc in the
plane of the sky. 

Once we obtained the 1.1 mm and 22 \mic\ fluxes from the photometry, 
\sh\ and \ssfr\ were calculated using Equation \ref{eq:sh} and \ref{eq:ssfr}
respectively. All the data points for the 10\arcmin, 15\arcmin, and
20\arcmin\ resolution are above the detection limit in both \sh\ and \ssfr.
The correlation between log(\sh) and log(\ssfr) increases from a linear 
correlation coefficient of 0.66 at 10\arcmin\ scale to 0.76 at 20\arcmin\ scale 
(Table~\ref{co}). 
Two methods of linear curve fit were performed on the data: 
an unweighted least-squares fit using MPCURVEFIT \citep{markwardt09} and 
a robust bisector linear fit (IDL Robust\_Linefit).
The relations for the convolved scales of 10\arcmin, 15\arcmin, 
and 20\arcmin\ can be seen  
in Figure~\ref{fig:sfl}(a). Each data point corresponds to a region inside an aperture centered on a 
peak of emission in the 1.1 mm image. The solid black line represents the 
robust fit to the data, the dotted grey line represents the least-square 
fit to the data, the dashed red line represents the extragalactic 
star formation relation (Equation \ref{eq:ks}), and the 
dot-dashed blue line represents the dense gas relation from \citet{wu05}. 
The coefficients of the curve fits are shown in Table~\ref{co}. 

The star formation relations are slightly sub-linear at all scales, more
so for the least-squares fit.  Both methods show 
increases in the intercept $a$ (the effective star formation rate) 
as the scale gets larger. 

\subsection{Analysis based on 22 \mic\  Peaks}\label{irpeaks}

\begin{figure}[h]
\center
\includegraphics[scale=0.6]{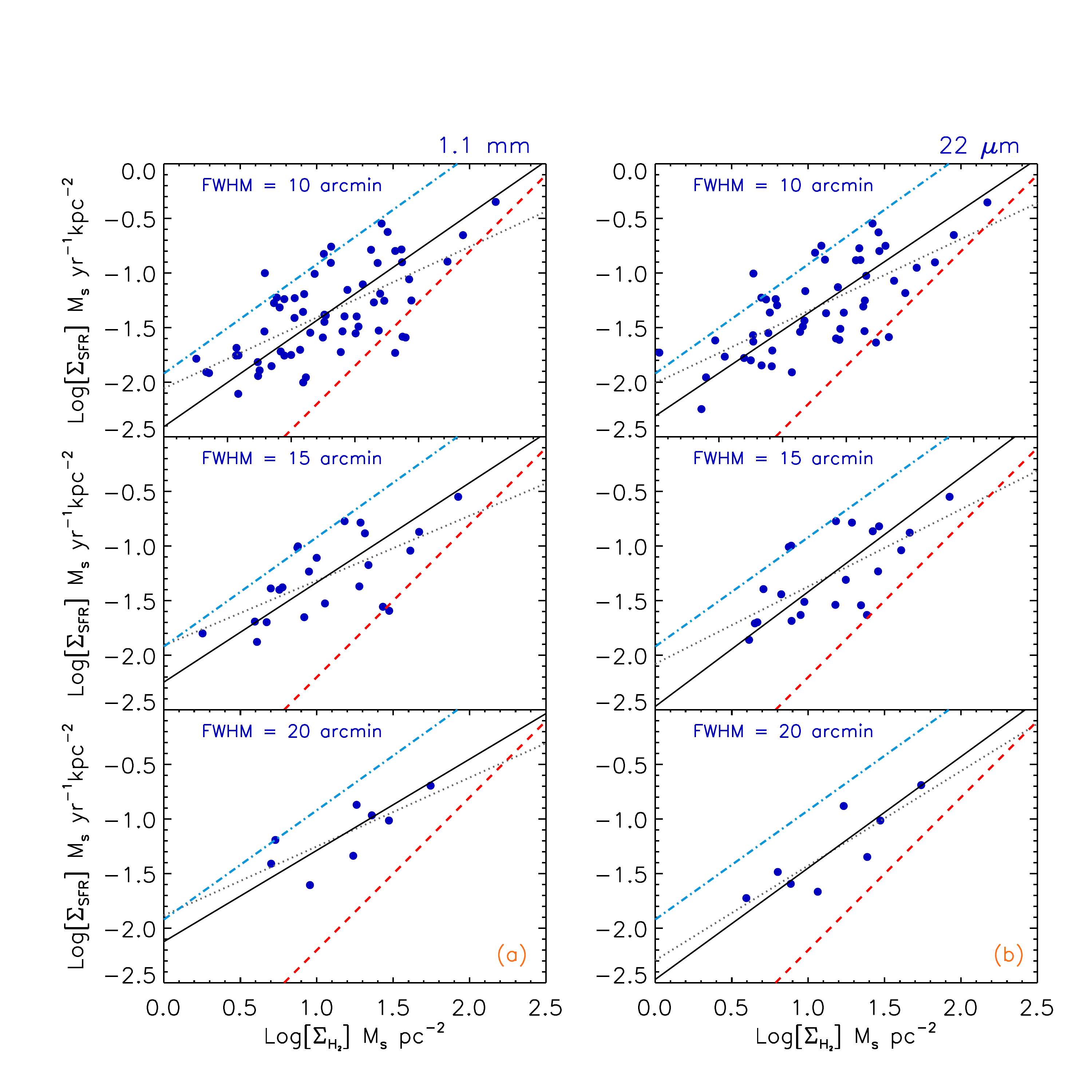}
\caption{\ssfr\ versus \sh\ from aperture photometry centered at 1.1 mm peaks (a) 
and 22 \mic\ peaks (b)
for image resolutions of 10\arcmin, 15\arcmin, and 20\arcmin.
The solid black line in each plot 
represents a robust linear fit to the data while the dotted grey line
represents a least-square fit. 
The dashed red line represents the extragalactic relation from Kennicutt et al. (1998)
and the dot-dashed blue line is the dense gas relation from Wu et al. (2005).}
\label{fig:sfl}
\end{figure}

What changes if
regions are identified using 22 \mic\ peaks instead of 1.1 mm peaks?
To answer that question, we used procedures similar to those used in the 1.1 mm peaks analysis, 
but we started by identifying local emission peaks in 22 \mic\ images 
at the resolution of 10\arcmin, 15\arcmin, and 20\arcmin\
and performed photometry on both 1.1 mm and 22 \mic\ images at 
locations of the 22 \mic\ peaks. 
Figure~\ref{fig:sfl}(b) shows the plot of the star formation relation 
at the three resolution scales. 
The fit parameters are included in Table~\ref{co}. 
The fit parameters are comparable to the parameters for the 1.1 mm
peaks.

Initial analysis did not show a significant difference when choosing 22 \mic\ 
peaks versus choosing 1.1 mm peaks as can be seen from the values of 
\tdep\ in Table~\ref{co}. 
When thresholds were applied to the data however, the result was different.
Instead of choosing all the identified local peaks, we chose only
bright local peaks by dropping all the regions with fluxes less than 
100 signal-to-noise ratio. The approximate values corresponding to
these thresholds are \sh\ $\approx$ 15 \msun\ pc$^{-2}$ and 
\ssfr\ $\approx$ 0.1 \msun\ yr$^{-1}$ kpc$^{-2}$. 
After the cut, the number of sources are 24, 11, and 6 for the 1.1 mm peaks and 
14, 9, and 3 for the 22 \mic\ peaks for the scales of 10\arcmin, 15\arcmin, and 20\arcmin\ 
respectively.
Regions with 1.1 mm peaks and 22 \mic\ peaks with fluxes above the 
threshold were then compared. 
The result in Figure~\ref{fig:irmm} shows a larger \tdep\ for 1.1 mm 
peaks than for 22 \mic\ peaks. 
The difference in \tdep\ decreases as the scale increases. 
The difference in log(\tdep) is about 0.37 at 5\arcmin\ scale and goes
down to about 0.05 at 20\arcmin\ scale.
 Note, however, that the scatter $\sigma(\log(\tdep))$ is about 0.4 at
 5\arcmin\ scale and about 0.2 at 20\arcmin\ scale.

\section{Discussion}\label{disc}

\subsection{Relating to Physical Scales}\label{physscales}

\begin{figure}[h]
\center
\includegraphics[scale=0.8]{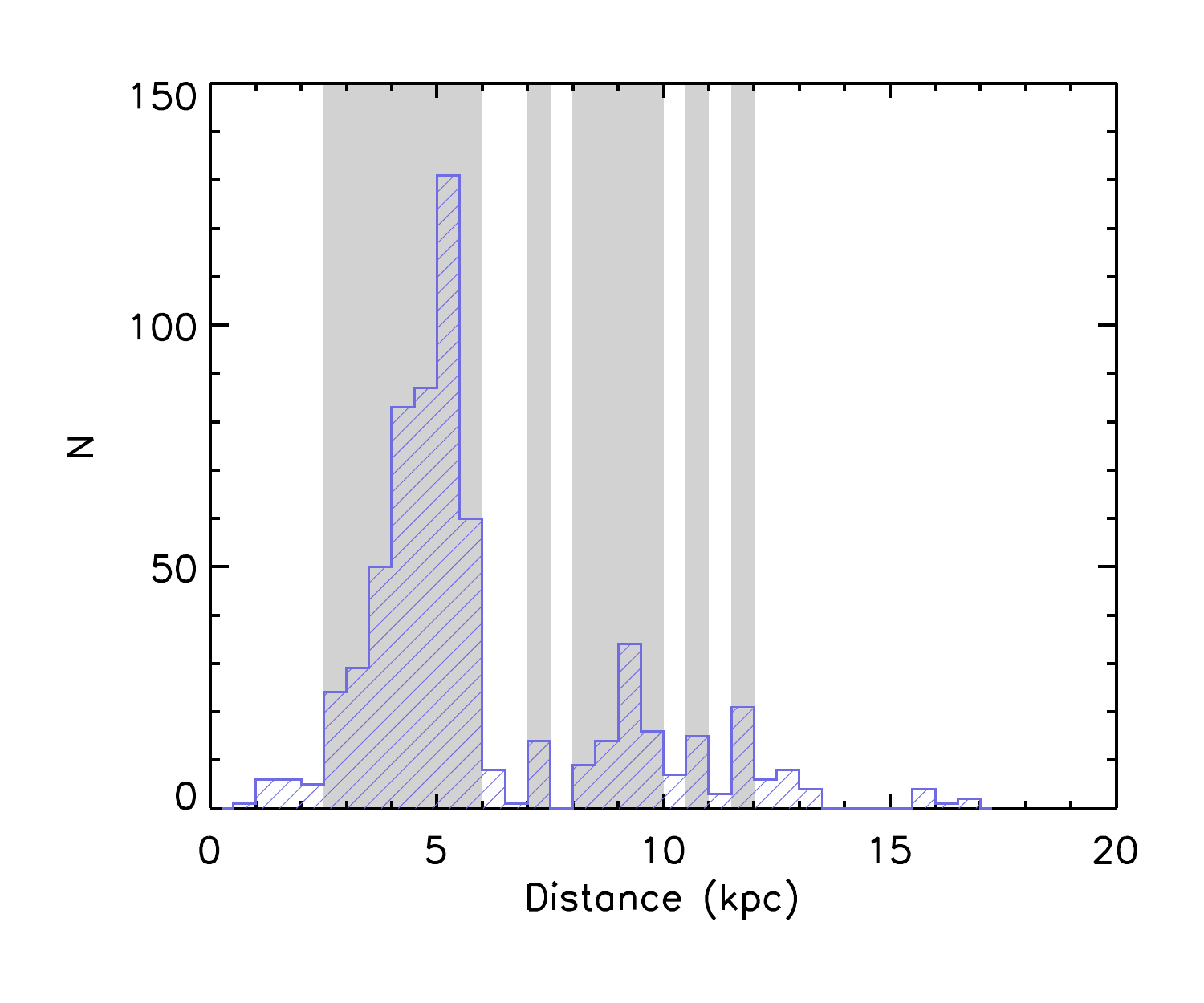}
\caption{The distribution of the distances
to a subset of the Bolocat sources inside our target regions 
from Ellsworth-Bowers et al. (2014) 
shows a large fraction of the sources locate around 5 kpc, 
and a smaller fraction of the sources locate around 10-12 kpc.
The mean of the distances is 5.9 kpc, and the median is 5.1 kpc. 
The shaded grey area represents the area inside which the number of sources 
accounts for 90\% of the total sources. }
\label{fig:dis}
\end{figure}

We have been looking at the relations between gas and star formation
at various angular scales. 
Converting the angular scales to physical scales for our Galactic
plane data is different than for resolved extragalactic data since
we are looking edge-on through the disk instead of face-on or nearly
face-on. However, the available distances to a subset of the sources
can be used to obtain some rough estimations of the physical size that
corresponds to each angular resultion. About 33\% of the Bolocat
sources in our targeted regions
have distances measured with a distribution shown in Figure~\ref{fig:dis}
\citep{ellsworth14}. 
In estimating the physical scales, 
we assumed that the distance distribution of these sources is representative
of the whole sample (see \S 5.1).

\begin{figure}[h]
\center
\includegraphics[scale=0.7]{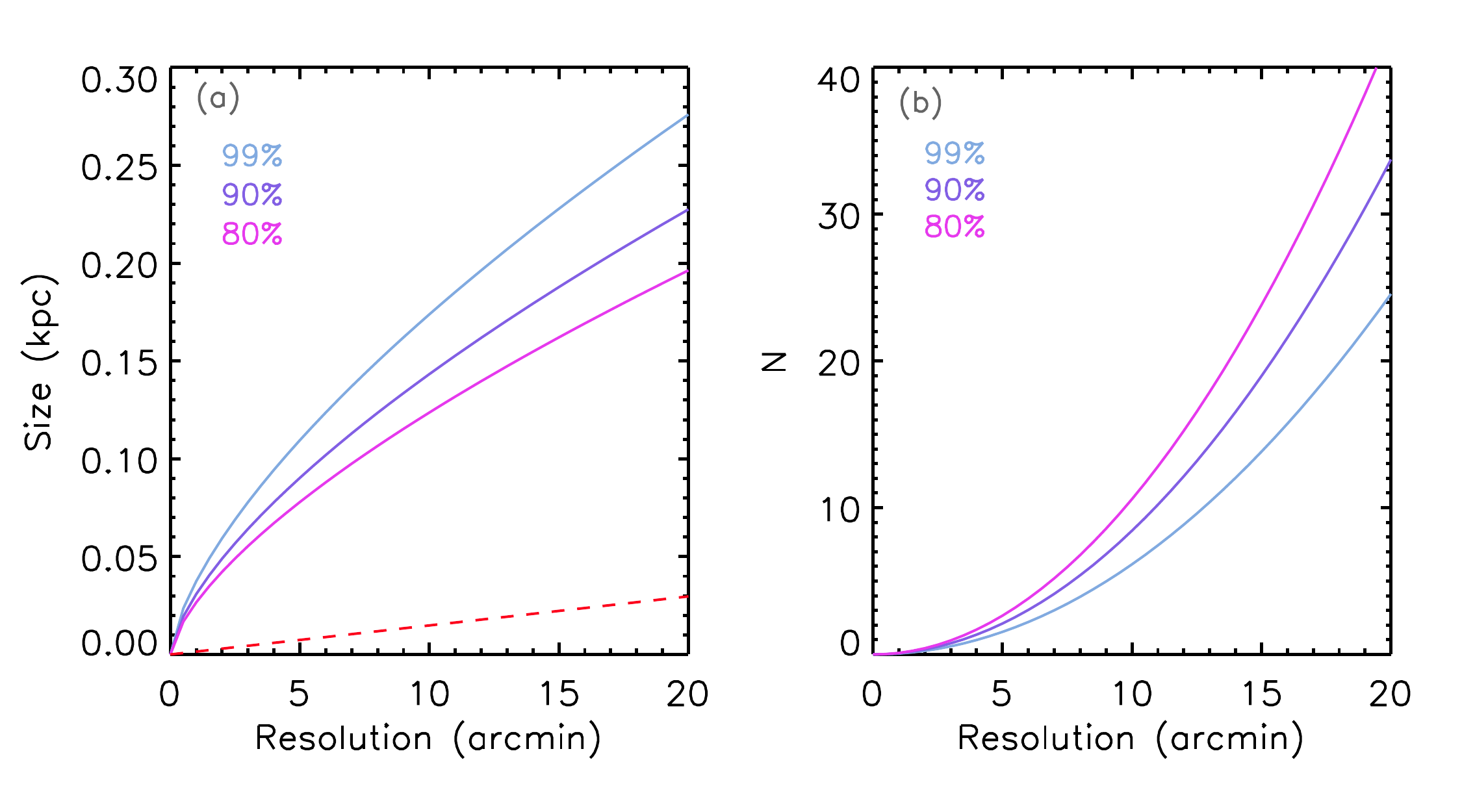}
\caption{(a) Effective averaging scales as a function of angular resolution
estimated by using a median distance of 5.1 kpc and distance ranges 
corresponding to source percentage of 80\%, 90\%, and 99\% of total sources. 
The dashed red line shows the transverse physical sizes at a distance
of 5.1 kpc.
(b) Estimated number of 1.1 mm sources inside one resolution element
for areas covering source percentage of 80\%, 90\%, and 99 \% of total sources. }
\label{fig:size}
\end{figure}

The first estimation of the physical scale used the  
median distance to the sources to calculate the size in the plane
of the sky. With a median distance of 5.1 kpc, physical sizes in the
plane of the sky at different resolutions are shown as the red, dashed
line in Figure~\ref{fig:size}(a). These sizes range from those of
clumps to those of clouds.
For angular scales small enough to be comparable to an individual source
size, the estimated physical scales should represent the physical
sizes of the sources. 

For comparison to extragalactic measurements, we also computed the
typical averaging scale for each
resolution, using the distance distribution.
The idea behind the method is to use source locations in Galactic latitude and
longitude to estimate the solid angle the sources subtend and
use the distance distribution to estimate the range of distances for
the sources. 
The result gives the volume subtended by the sources, which can be
converted to a length scale.
This would represent an upper limit to the relevant
scale.
To do this, first we binned the distance distribution with a binning
size of 0.5 kpc as shown in Figure~\ref{fig:dis}. 
Three values were chosen as source number density thresholds (per bin)
so that the total number of sources above the thresholds
account for about 80\%, 90\%, and 99\% of the total sources.
This is analogous to drawing number density contours on an image; however,
instead of an image, we are drawing one-dimensional contours on a
distribution.  
The shaded-grey area in Figure~\ref{fig:dis} is an example of the area
inside the contour of 90\% of the total sources.
The resulting distance ranges ($\delta d$) are 4.5, 7, and 14 kpc for 
80\%, 90\%, and 99\% of the total sources respectively. 
We calculated an averaging volume ($V$) for the very long, skinny rectangular
prism from 
\begin{equation}
\frac{V}{\text{kpc}^3} = \left( \frac{\theta_{r}}{\text{rad}} \frac{d}{\text{kpc}} 
\right)^2 \times \frac{\delta d}{\text{kpc}},  
\end{equation}
where $\theta_r$ is the angular size of a resolution element.
Reduced to a sphere, the effective radius $R$ is 
\begin{equation}
\frac{R}{\text{kpc}} = \left( \frac{3}{4\pi} \frac{V}{\text{kpc}^3}
\right)^{1/3}. 
\end{equation}
The effective averaging scale, the diameter ($2R$), at different resolutions is shown in
Figure~\ref{fig:size}(a). Using this estimation at the median 
distance $d$ of 5.1 kpc, the angular resolution
of 20\arcmin\ corresponds to an averaging scale of $\approx 225$ pc.

A similar method was used to estimate the number of Bolocat sources
each resolution element contains. Bolocat sources were mapped in the Galactic
coordinates, and a source number density map was created by counting number of 
sources inside bin sizes of 0.2$^\circ$ in both axes.
Then 2D contours of source number density 
were drawn on the source number density map to contain approximately 80\%, 90\%, and 99\% of
the total number of sources. The total solid angle
inside each contour gave a total area for each completeness value. 
The average number of sources per resolution element is the solid angle of the
resolution element ($\theta_{r}^2$) divided by the total area for the completeness value 
times the total number of sources. 
The average number of sources per resolution element is shown in
Figure~\ref{fig:size}(b). For the larger smoothing lengths, we
are typically averaging over 20 to 40 sources.

\subsection{The Scatter in the Star Formation Relation}\label{scatsfr}

\begin{figure}[h]
\center
\includegraphics[scale=0.7]{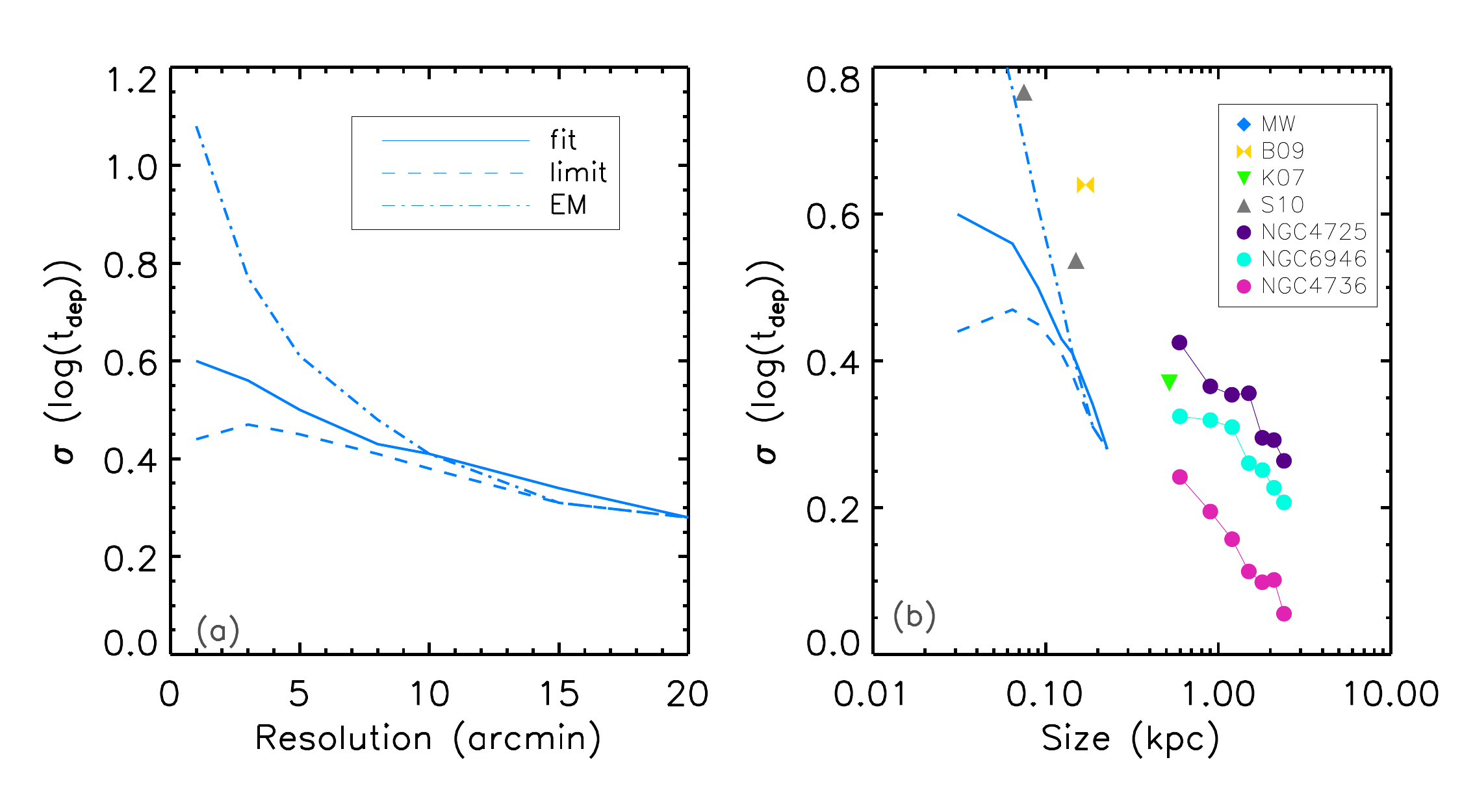}
\caption{(a) The standard
deviation of the log( \tdep) at different scales, calculated by
three different methods for the pixel-by-pixel analysis.
$fit$ shows the result from a Gaussian fit to the distribution of log(\tdep) for all data points with positive flux. 
$limit$ refers to the standard deviation of the log(\tdep) after upper limits were assigned to \sh\ and \ssfr.
$EM$ refers to the standard deviation from the expectation-maximization method, which estimated the censored data assuming a normal distribution of log(\tdep). 
 (b) Comparisons of the standard deviation of the log(\tdep) 
between the Galactic Plane from this study and the extragalactic
studies of M51(a) by 
Blanc et al. 2009 (B09), NGC5194 by Kennicutt et al. 2007 (K07), M33
by Schruba et al. 2010 (S10), 
and Leroy et al. 2013 (NGC4725, NGC6946, NGC4736). All the data for
extragalactic studies were taken from Leroy et al. (2013)}
\label{fig:scatter}
\end{figure}

The results on the star formation relations show that the 
relations at small scales have large scatter as seen in the 
Bolocat source case or at the small scales in the pixel-by-pixel analysis. 
The rank correlation coefficients between log(\sh) and log(\ssfr) increase
from 0.40 at the Bolocat source scale to 0.79 at 20\arcmin\ scale for
sources based on 1.1 mm peaks. 
This result is clearly seen in the
pixel-by-pixel star formation relation and the distribution of the \tdep. 
The scatter of the \tdep\ in the pixel-by-pixel analysis was estimated by three methods
(Table~\ref{tdep}), and the results show decreases of
$\sigma(\log{\tdep})$ as the scale gets larger. 
Figure~\ref{fig:scatter}(a) shows the scatter $\sigma(\log{\tdep})$
over resolution scale for the three methods. 
At small scales (1\arcmin\ -3\arcmin), the EM method gives much
larger values of the scatter than the other two methods. At these
scales, over 50\% of the data are below detection limits, making 
the estimates of $\sigma(\log{\tdep})$ uncertain. 
The three methods give comparable values for scales over 8\arcmin.

Figure~\ref{fig:scatter}(b) shows a comparison of our results with some resolved 
extragalactic star formation relations. We used the averaging scale
with the 90\% distance contour as the maximum relevant scale.
We caution that we are comparing across different data sets, and the 
differences in observations, methodologies, and other factors 
could contribute in the differences in the scatter of \tdep. 
Our result for the Galactic Plane covers small scales where only few
comparable extragalactic data exists to a scale of about 200 pc. 
The trend from our data suggests a smaller scatter in the depletion
time for the Galactic Plane than the extragalactic data. 

One important difference in our study is the choice of molecular gas tracers. 
The BGPS 1.1 mm in general traces denser and smaller parts of molecular 
clouds than CO or $^{13}$CO \citep{battisti14}. 
A single GMC can contain multiple 1.1 mm sources. The smaller scatter 
in our result is consistent with the fact that star formation is 
more closely associated with denser regions than with the general 
molecular cloud.

To examine possible causes of the change in the scatter, we looked at 
contributions to the scatter in log \tdep\ ($\sigma(\log{\tdep})$). 
Several possible sources of uncertainties in log \tdep\ include
observational and photometric uncertainties, 
uncertainties in the parameters assumed in calculating \ssfr\ and \sh\ ,
 uncertainties in the 22 \mic\ flux due to spatial
offsets between 1.1mm and 22 \mic\ emission, 
and a scatter due to variations in intrinsic properties of the
sources.

The uncertainties from the observations and the photometry
were estimated as the errors associated with the data. 
The other sources of uncertainties will be discussed below.  

\subsubsection{Parameter Uncertainties}
In the calculation of \sh, we assumed a dust temperature of 20 K,
a gas to dust mass ratio of 100, and 
a dust opacity of 1.14 cm$^2$ g$^{-1}$
 for all the sources following the 
study of \citet{schlingman11}. The variations in the 
real values would contribute to the scatter in \sh. 
Spectroscopic observations of several molecular lines for 
Bolocat sources show a median temperature of $\approx 18$ K
with a temperature range from 10-30K \citep{shirley13}. 

In calculating \ssfr, we used the SFR - 24 \mic\ relation (Equation \ref{eq:ssfr}) from 
\citet{calzetti07}. The calibration was derived assuming a constant SFR on a timescale of 
100 Myr and a Kroupa IMF. When applied to an individual molecular cloud
or a star forming region, the assumptions cannot always be valid. 
The timescale assumed for constant star formation is much longer than 
the average lifetime of molecular clouds \citep{murray11}. 
Several studies show that infrared tracers 
underestimate SFR with large uncertainties in clouds with low mass or
low SFR \citep{vutisalchavakul13, dasilva12, dasilva14}.
The combined effects of stochastically sampling the IMF and
the star formation history causes SFR indicators such as H$\alpha$, 
FUV, and bolometric luminosity to under-estimate the SFR \citep{dasilva14},
 and the size of the underestimate gets larger as the SFR gets smaller. 
\citet{dasilva14} also showed that when stochasticity is taken
into account, SFR indicators do not provide a unique value of the SFR. 
 
The variations in the properties of 
each source contribute to the uncertainties in how well 
the calculated SFR agrees with the true SFR. 
This is especially important for the sources with low \ssfr\ in our
data.
As discussed earlier in \S \ref{iremission}, infrared luminosity 
starts to underestimate SFR below $10^{-5}$ \msun\ yr$^{-1}$ \citep{dasilva14}. 
Using the averaging scales with a distance contour of 90\%, we estimated the corresponding 
\ssfr\ by assuming a projected area of $\pi R^2$.
The value of SFR translates to 
\ssfr\ $\approx 1.4 \times 10^{-2}$ \msun\ yr$^{-1}$ at 1\arcmin\ scale, $3.1 \times 10^{-3}$ 
 \msun\ yr$^{-1}$ at 3\arcmin\ scale, 
and $1.6 \times 10^{-3}$  \msun\ yr$^{-1}$ at 5\arcmin\ scale. 
Comparing these values to our results from the pixel-by-pixel analysis (Figure~\ref{fig:pixsfl}) 
shows that some fractions of the data points at 
1\arcmin- 3\arcmin\ scales are below the values, so they are affected by the bias in \ssfr. 
Above 5\arcmin\ scale, all the data points are above the bias values for both pixel-by-pixel and 
source based results.

The method of choosing regions also affects
the uncertainty introduced by IR as a SFR tracer.
Pixel-by-pixel analysis is affected the most since the data contain 
regions with low \sh\ and low \ssfr. 
Regions chosen by identifying bright emission peaks will show less variations 
since these regions were chosen based on assumption of strong emission.
Therefore, SFR measurements from infrared tracers are more reliable when applied to 
regions with strong IR emission peaks as in Figure~\ref{fig:irmm}. 
For extragalactic studies, this effect will be more important in 
line-of-sight (pixel-by-pixel) studies than in studies with CO, H$\alpha$, or IR peaks. 

\subsubsection{Spatial offsets between IR and 1.1 mm}\label{offset}
The 1.1 mm emission comes from cold dust from dense molecular 
gas regions in GMCs while the 22 \mic\ emission should be dominated by
warmer dust heated by stellar radiation. The two emitting regions
might not perfectly coincide spatially with each other. 
The sizes of the emitting regions could also be different for 1.1 mm 
and 22 \mic. These two factors will result in spatial offsets,
contributing to the scatter in the star formation relation if the 
scale size is smaller than a typical offset. 

\begin{figure}[h]
\center
\includegraphics[scale=0.8]{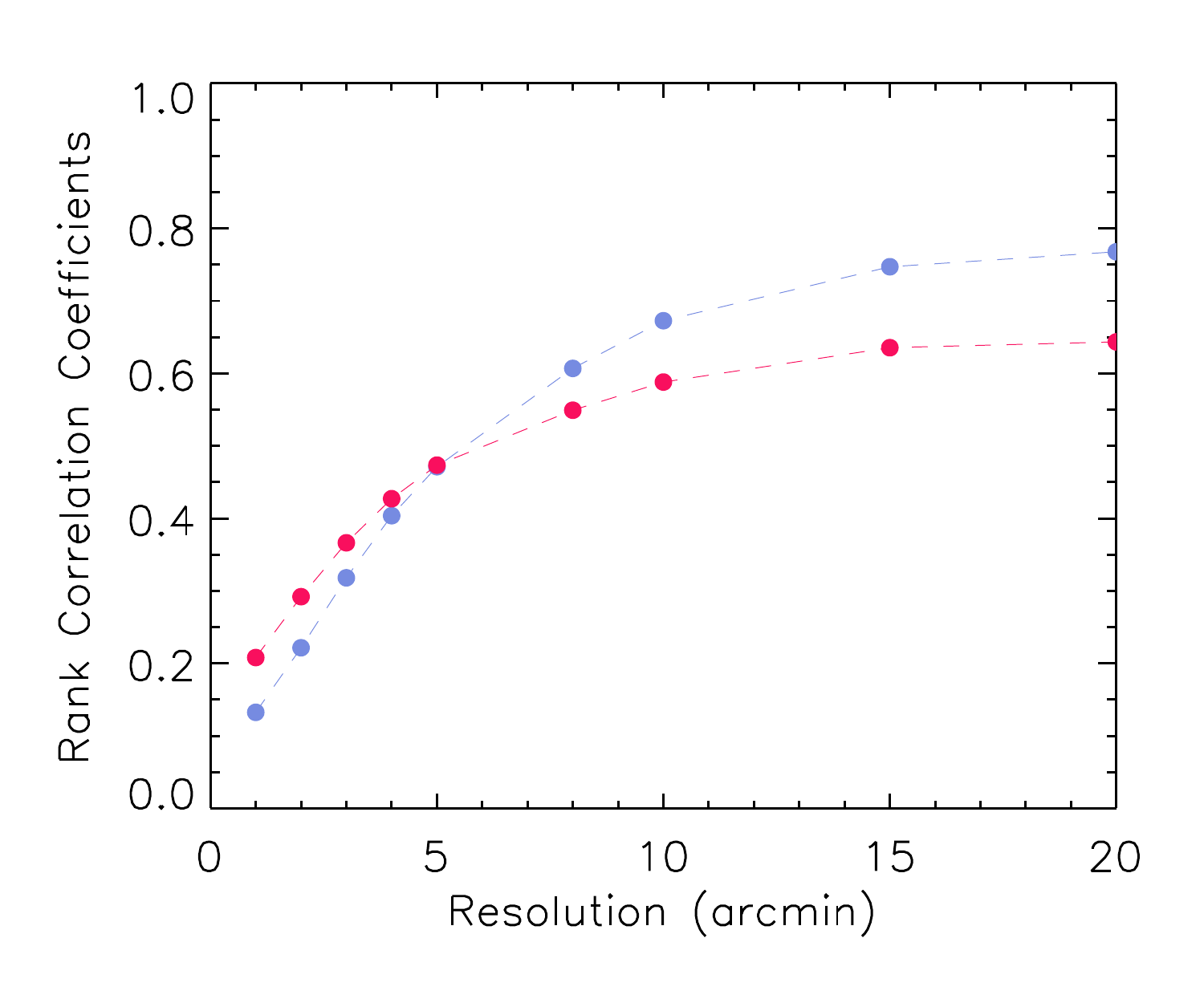}
\caption{The pixel-by-pixel rank correlation coefficients 
between 1.1 mm images and 22 \mic\ images
increase steeply at small scales and asymptote around 5-8\arcmin\
for both regions. The blue data points 
represent the correlation coefficients from region 1 
, and the red data points represent region 2.}
\label{fig:pixcor}
\end{figure}

If there is a general offset between the 22 \mic\ and 1.1 mm sources, 
the pixel-by-pixel correlation between the two images should get 
better once the spatial resolution becomes larger than the offset. 
From Figure~\ref{fig:pixcor}, the correlation increases 
rapidly at small scales until they level off at around 5-8\arcmin\
scale. This result suggests that there is an offset of small 
scale variations between 22 \mic\ and 
1.1 mm emission of about  5-8\arcmin, corresponding to 7 to 12 pc
at the median distance of 5.1 kpc. This offset is typical of cloud
sizes, consistent with the idea that the 1.1 mm source may be a
remnant clump, while the infrared emission traces star formation
in a now-destroyed clump in the same cloud.

For the case of the Bolocat sources, \tdep\ from the data 
is greater than the average \tdep\ at larger scales. 
The \tdep\ of $\approx 1$ Gyr is close to the average 
values found in extragalactic studies \citep{wong02, leroy13}. 
The 22 \mic\ flux for each 1.1 mm source was calculated by 
centering the photometry aperture on the 
center of the 1.1 mm source. If the infrared emission 
associated with the 1.1 mm sources does not coincide with 
the 1.1 mm peak then the estimated 22 \mic\ flux would not 
be representative of the total emission.
The infrared emission could also be more extended than
the size of the aperture used in the photometry, in which case we
would be underestimating the SFR. 

To investigate this issue, we looked at the 22 \mic\ emission 
for several sources from the images. One of the sources we tested was
 G23.95+0.16, which is a massive dense clump with an observed water
 maser. This source had previously been studied by \citet{wu10,
 vutisalchavakul13}. The source size obtained by 
fitting a 1D Gaussian is about 3.9\arcmin, much larger than our
aperture size of 80\arcsec. A large fraction of the IR emission lies
outside the aperture resulting in an underestimated SFR. 
However this particular source has high \sh\ and \ssfr\ compared to the 
whole sample. We looked at several other 1.1 mm sources for which
there are associated 22 \mic\ emission and found that most of the 
22 \mic\ emission is more extended than the aperture size. 
To see how much this issue affected the star formation relation 
result, we performed the photometry again with a larger aperture size
of 160\arcsec. The result shows a higher average \ssfr, and the 
relation now lies above the \citet{kennicutt98} relation.

\subsubsection{Intrinsic Source Properties}

Aside from the uncertainties already discussed, the 
scatter in the star formation relation can also be 
contributed by intrinsic variations in the relation itself. 
If the SFR for each source is not determined only
by the amount of gas available, then we would not 
expect to see a tight correlation between the two. 
Our results show a much larger scatter at small scales
than do the relations found in disk-average studies. 
What causes the difference? 
These 1.1 mm sources are expected to be star forming regions. 
Then these star forming regions should have variations in 
their properties. The sources that are in earlier stages of star
formation would contain a large amount of gas and little 
infrared emission from star formation (large \tdep), while the sources that 
are in later stages of star formation would contain less gas due to
gas depletions and emit more strongly in the IR due to stars and current
star formation (small \tdep). Battersby et al. (2010) found some of the 
1.1 mm sources to be infrared dark clouds (IRDC).
They are in an early stage of star formation, which would show
little or no SFR. 
If the 1.1 mm sources are in different evolutionary stages, then 
they will show a large scatter in the star formation relation at 
the scale of individual sources.
Sampling a larger number of individual regions at different
stages averages out the scatter in \tdep, resulting in  
a decrease in the scatter in the relation. 

The effect of sampling different stages of star formation  on the
star formation relation has been a topic of several recent studies 
\citep{onodera10, schruba10, leroy13, kruijssen14}. 
\citet{schruba10} showed, in a study of star formation in 
M33 at the scale of 75 to 1200 pc, 
that the choice of CO or H$\alpha$ peaks as centers gave different values
of \tdep.
 The difference between the \tdep\ from CO and H$\alpha$
peaks decreases as the aperture size increases. They argued that the 
dependence of \tdep\ on scale was mainly due to the effect of sampling 
different evolutionary stages. 
\citet{kruijssen14} constucted a model to describe the
dependence of \tdep\ and the scatter in \tdep\ on spatial scales
and how the differences between \tdep\ when choosing regions on 
either gas or star formation tracer peaks can be used to estimate
the timescales involved in star formation processes.

We found similar results 
when comparing regions centered on 1.1 mm peaks with 
regions centered on 22 \mic\ peaks. 
Regions with strong 1.1 mm emission are more likely to 
be at an earlier stage of star formation where there is 
still a large amount of gas while regions with strong 
22 \mic\ have already formed stars.
The differences in the average \tdep\ between choosing 
different peaks shown in Figure~\ref{fig:irmm} 
support the hypothesis that the differences in the stage of star formation
contribute to the large scatter in the star formation 
relation at small scales. 
\\

All the mentioned sources of scatter can affect the relations 
between \ssfr\ and \sh. 
To quantitatively explain the observed data with these 
uncertainties requires a careful modeling of how each 
source of scatter depends on scale, which is beyond the scope 
of this paper. Future studies of properties of individual 
star forming regions, especially their evolutionary stages,
will provide more insights into the problem. 

\subsection{The Star Formation Relation and Depletion Times}

Aside from looking at the scatter in the relation, we can also look
for changes in the form of the star formation relation as the resolution
is changed. While the correlation improves with averaging
scale, the changes to the fitted values for the slope and intercept 
are not very significant.
The  star formation relations at all scales 
are slightly sub-linear and lie between the extragalactic relation from 
\citet{kennicutt98} and the dense gas relation of \citet{wu05}.

The typical depletion time, both from the pixel-by-pixel analysis
and from the source-based analysis with large averaging scales is
200 Myr. 
The Bolocat source-based analysis shows a larger \tdep\ of about 1 Gyr,
closer to the typical extragalactic value. However, this value is very
likely an overestimation. 
As discussed in \S \ref{offset}, the \ssfr\ for the Bolocat sources
are underestimated due to both the fact that not all the infrared
emission was inside the aperture and the bias from using 22 \mic\ to
trace SFR for low mass sources. For this analysis we also centered
the apertures on Bolocat source, which are bright 1.1 mm regions, so
the analysis biases toward larger \tdep.

\begin{figure}[h]
\center
\includegraphics[scale=1]{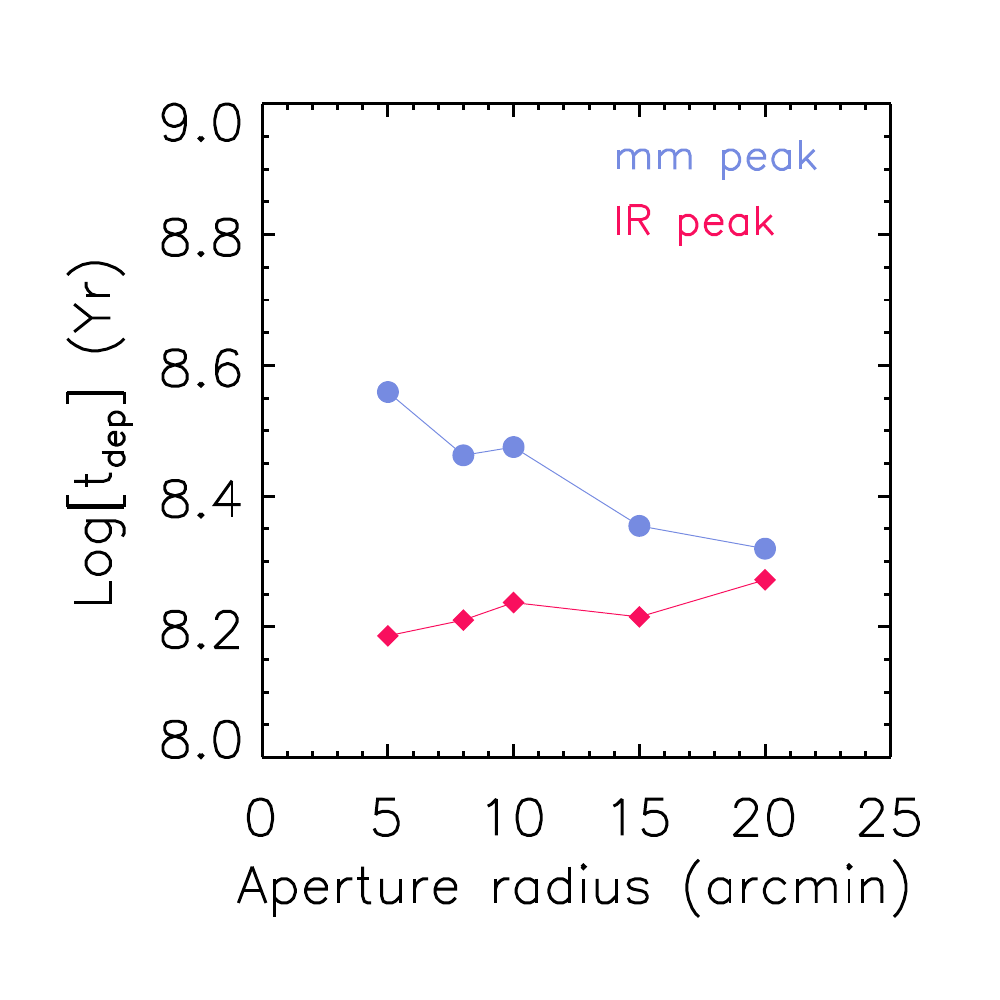}
\caption{Comparing the average depletion time between two methods of
  selecting regions show that the avereage \tdep\ is larger for regions centered at 1.1 mm
  peaks than for regions centered at 22 \mic\ peaks when cuts of 100
  signal-to-noise ratio were applied to \sh\ and \ssfr. 
}
\label{fig:irmm}
\end{figure}

The pixel-by-pixel analysis does not show variations 
in the average \tdep\ over resolution scales. 
When all the regions are sampled,
\tdep\ can be described reasonably well with log-normal
distributions centered at $\approx 10^{8.3}$ yr. 
Choosing bright 22 \mic\ regions is equivalent to sampling the lower
tail of the distribution of \tdep, while choosing bright 1.1 mm regions is
equivalent to sampling the higher tail of the distribution, resulting
in the differences in \tdep\ as seen in Figure~\ref{fig:irmm}. 
Evidently, the method of choosing regions affects the result of \tdep, 
as does the method of identifying local emission peaks. 
Before making the cut in the 22 \mic\ and 1.1 mm flux, the data did
not show a clear trend in \tdep\ over spatial scale. 
This is likely due to the fact that without the cut, all the
identified regions were included. The lower brightness regions tend to
sample near the center of the \tdep\ distribution, therefore
lowering the distinction between IR or mm peaks.  
Data for other galaxies could be affected as well since the
sensitivity limit varies between data sets.

The constant timescale of 200 Myr seen throughout our data set 
is similar to the mean value found in the nearby
clouds, but about 5 times greater than that found for
the dense gas ($A_V > 8$ mag) in the nearby clouds \citep{evans14}. 
Since the 1.1 mm emission is mostly tracing clumps with $\mean{n} \approx
10^{3.5}$ \cmv, similar to the mean density within the $A_V > 8$ mag
contours \citep{evans14}, this difference may indicate a systematic
underestimate of the star formation rate from the 22 \mic\ emission.
Individual YSOs could be counted in the nearby clouds, rather than
relying on the 22 \mic\ emission. \citet{vutisalchavakul13} showed
that the mid-infrared emission does underestimate star formation rate
in the nearby clouds where high-mass stars are rare. For this reason
we believe that the actual value of \tdep\ is likely overestimated.

On the other hand, the likely overestimated value we get for \tdep\ 
is already 5 times {\it smaller} than that found in other galaxies. 
The BGPS 1.1 mm emission we used in this study traces denser gas than the 
common tracers used for other galaxies. The 1.1 mm emission only traces about 11\% 
of the gas traced by $^{13}$CO, a more typical gas tracer in extragalactic studies 
\citep{battisti14}. Therefore, we would expect the average \tdep\ to be smaller 
than extragalactic values.
As a result of these systematic issues, these data tend to lie between
the Kennicutt relations for total gas and the Wu relation for even 
denser ($\mean{n} = 10^{4.5}$ \cmv) gas \citep{wu05}.

\section{Summary}\label{summary}
We studied the relationship between molecular gas and SFR surface density 
for 11 deg$^2$ of the Galactic Plane. 
The 1.1 mm data from the Bolocam Galactic Plane Survey, 
which traces dense gas inside molecular clouds, 
was used as a tracer of molecular gas while 22 \mic\ data 
from the WISE All-Sky survey was used to trace SFR. 
We studied the relation from the scale of 33\arcsec\ to the largest 
scale of 20\arcmin\ by convolving images with Gaussian beams.   
We started by looking at the correlations between 22 \mic\ and 1.1 mm images pixel-by-pixel 
and found that the rank correlation coefficient increases rapidly 
as scale size increases, 
leveling off at the scale of about 5\arcmin-8\arcmin. 
The 22 \mic\ and 1.1 mm 
emission are already well correlated at the 
scale of 5\arcmin\ and the correlations
do not change much at larger scales, 
suggesting a spatial offset or small scale variations around this
scale, which corresponds to estimated physical scales of 7-12 pc. 

We studied the star formation relations both by analyzing pixel-by-pixel values
and by identifying 1.1 mm and 22 \mic\ peaks.
The star formation relations from the pixel-by-pixel analysis show close to linear relations.
The distribution of \tdep\ can be closely represented as a log-normal 
with the avereage of about 200 Myr regardless of the resolution. 
The relation on small scales shows large scatter, and the scatter decreases as the scale gets larger. 
The scatter of the log(\tdep) decreases from above 0.6 at 1\arcmin\ scale 
to about 0.28 at 20\arcmin\ scale. 
The typical depletion times lie between those for dense clumps and those
for total gas or for molecular gas in other galaxies.

For sources centered at 1.1 mm peaks, we found a weak correlation between 
\ssfr\ and \sh\ at 1.1 mm source scale (aperture diameter of 80\arcsec).
The correlation gets better at larger scales similarly for 1.1 mm
peaks and 22 \mic\ peaks. There are no significant differences in the form of 
the relation at different scales or when comparing 1.1 mm to 22 \mic\ peaks.
The star formation relations at all scales are slightly
sublinear and lie above the
extragalactic relation from \citet{kennicutt98}. 
When selecting only bright peaks however, the average \tdep\ 
from centering at 1.1 mm peaks is larger than the average \tdep\ 
from centering at 22 \mic\ peaks. 
These differences in \tdep\ decrease as the scale increases.

The average depletion time of 200 Myr seen across the data is 
about 5 times smaller than the typical \tdep\  of 1 Gyr in extragalactic studies
and larger than the \tdep\ measured for dense clumps. 
The smaller depletion time for the Galactic Plane than the extragalactic value
can be explained by the fact that the 1.1 mm emission used as a gas
 tracer for this study traces denser gas than the usual gas tracer such as $^{12}$CO or $^{13}$CO. 
The 22 \micron\ emission could also be systematically underestimating the SFR across all scales. 
\\

We thank Erik Rosolowsky and the rest of the  BGPS team for useful 
discussions and assistance. 
We thank the referee, Andreas Schruba, for constructive comments that helped improve the
paper. This research made use of Montage, 
funded by the National Aeronautics and Space Administration's 
Earth Science Technology Office, Computation Technologies Project, 
under Cooperative Agreement Number NCC5-626 between NASA and the 
California Institute of Technology. 
Montage is maintained by the NASA/IPAC Infrared Science Archive.
This publication makes use of data products from the Wide-field Infrared Survey Explorer, which is a joint project of the University of California, Los Angeles, and the Jet Propulsion Laboratory/California Institute of Technology, funded by the National Aeronautics and Space Administration. 
This work was supported by 
NSF Grant AST-1109116 to the University of Texas at Austin.

\clearpage

\begin{table}[h]
\center
\caption{The depletion time and the scatter in the depletion time at each resolution for pixel-by-pixel analysis}\label{tdep}
\begin{tabular}{l|lll|lll|r}
\hline
\multicolumn{1}{c|}{\multirow{2}{*}{Resolution(\arcmin)}} & \multicolumn{3}{c|}{$\mean{\log{\tdep}}$} & 
\multicolumn{3}{c|}{$\sigma(\log{\tdep})$} & \multicolumn{1}{c}{\multirow{2}{*}{$f$(limit)}} \\ \cline{2-7}
\multicolumn{1}{c|}{}                & fit        & limit        & EM       & fit          & limit          & EM         & \multicolumn{1}{r}{}  \\ \hline
 1                   & 8.49            & 8.30             & 8.07            & 0.60         & 1.00         & 1.84        & 0.52             \\
3                   & 8.30            & 8.30             & 8.30            & 0.56         & 0.75         & 1.06        & 0.35              \\
5                   & 8.30            & 8.33             & 8.36            & 0.50         & 0.60         & 0.73        & 0.22              \\
8                   & 8.33            & 8.34             & 8.36            & 0.43         & 0.48         & 0.53        & 0.11              \\
10                 & 8.34            & 8.34             & 8.34            & 0.41         & 0.41         & 0.42        & 0.05              \\
15                 & 8.34            & 8.34             & 8.34            & 0.34         & 0.31         & 0.31        & $<0.01$            \\
20                 & 8.37            & 8.37             & 8.34            & 0.28         & 0.28         & 0.28        & $<0.01$            \\ \hline
\end{tabular} \\
The average \mean{log(\tdep)} and the scatter of the log(\tdep) distribution
by three different methods. The $fit$ method estimates  
$\mean{\log(\tdep)}$ by fitting
a log-normal distribution to all data points with positive flux to
 get the mean and the standard deviation. 
The $limit$ method uses upper limit in \sh\ and \ssfr\ in calculating the median
of log(\tdep) and the standard deviation. 
The $EM$ method uses the expectation-maximization algorithm (Wolynetz 1979)
to estimate the mean and the standard deviation of censored 
normal distribution of log(\tdep). 
The last column, $f$(limit) gives the fraction of the data points with
an upper limit on either \sh\ or \ssfr. 
\end{table}

\begin{table}[h]
\center
\caption{Parameters for the source based analysis}\label{co}
\begin{tabular}{l|llllll}
\hline
\multirow{2}{*}{Resolution} & \multicolumn{6}{c}{1.1 mm}                   \\ \cline{2-7} 
                            & $\rho$ & $\rho$(rank) & robust(n, a) & LS(n, a)   & log(\tdep) & $\sigma(\log(\tdep))$ \\ \hline
10\arcmin                   & 0.66   & 0.63         & 0.97, -2.4   & 0.65, -2.1 & 8.43       & 0.34                  \\
15\arcmin                   & 0.65   & 0.62         & 0.91, -2.2   & 0.59, -1.9 & 8.34       & 0.32                  \\
20\arcmin                   & 0.76   & 0.79         & 0.84, -2.1   & 0.63, -1.9 & 8.32       & 0.24                  \\ \hline
                            & \multicolumn{6}{c}{22 \mic}                                                            \\ \cline{2-7} 
10\arcmin                   & 0.70   & 0.66         & 0.94, -2.3   & 0.66, -2.0 & 8.37       & 0.34                  \\
15\arcmin                   & 0.67   & 0.64         & 1.0, -2.5    & 0.71, -2.1 & 8.41       & 0.31                  \\
20\arcmin                   & 0.84   & 0.83         & 1.0, -2.5    & 0.87, -2.3 & 8.45       & 0.21                  \\ \hline
                            & \multicolumn{6}{c}{Bolocat}                                                            \\ \cline{2-7} 
33\arcsec                   & 0.50      & 0.40       & \multicolumn{2}{l}{0.84, -2.68$^a$}     & 9.20$^b$          & 0.59$^b$  \\ \hline
\end{tabular} \\
Note: (a) linear fit to the data using MPFITEXY (Marhwadt, 2009). \\
(b) Values correspend to mean and standard deviation of log(\tdep)
from the expectation-maximization (EM) method. 
\end{table}

\clearpage

\appendix

\section{Saturation}
WISE 22 \mic\ images contain some saturated regions with a large number of 
saturated pixels near peaks of bright, extended emission. 
The saturated areas are small compared to the total area, but saturation will 
affect larger portions of the image once we convolve to larger beam sizes. 
For further analysis, we replaced the saturated pixels with estimated values. 
Over an extended saturated region, the estimation was done by calculating average values of 
the surrounding regions and performing a thin plate spline interpolation. 
Consequently, the values of the flux over saturated regions have large uncertainties.
However, the saturated area only covers less than 0.01 percent of the entire image.
The uncertainties in the estimations should have minimal effect on the flux calculations. 

\section{Photometric Uncertainty} 
The uncertainties in the \ssfr\ for the Bolocat sources (\S \ref{bgpscat}) came from the 
estimated uncertainties in the photometry performed on WISE 22 \mic\ images. 
Since the photometry was performed directly from the WISE all-sky release images, the 
uncertainties were estimated following the Explanatory Supplement to the WISE 
All-Sky Data Release Products. The uncertainty of the source flux ($\sigma_{\text{src}}$)
was contributed by instrumental and calibrational uncertainties, Poisson noise, and 
uncertainties from background estimations. 
The $\sigma_{\text{src}}$ was estimated by: 
\begin{equation}
\sigma_{\text{src}}^2 = F_{\text{corr}} \left( \sum \sigma_{iA}^2 + \frac{{N_A}^2}{N_B} \sigma_{\text{B/pix}}^2 \right) ,
\nonumber
\end{equation}
where 
\begin{align}
F_{\text{corr}} &= \text{pixel to pixel correlated noise correction factor} \nonumber \\
\sigma_{iA} &= \text{flux uncertainty for each pixel inside the aperture}  \nonumber \\ 
\sigma_{\text{B/pix}} &= \text{uncertainty in the background per pixel}  \nonumber \\
N_A &= \text{number of pixels inside the aperture} \nonumber \\
N_B &= \text{number of pixels used to estimate the background flux}. \nonumber 
\end{align}
After obtaining the 22 \mic\ mosaics for the two regions, the images were convolved to a resolution 
of 33\arcsec\ to match the resolution of the 1.1 mm images from BGPS. $\sigma_{iA}$ were obtained
from the original uncertainty maps from the WISE all-sky release. The correction for correlated noise, $F_{corr}$, 
was obtained from the WISE Explanatory Supplement, for which it has been estimated for certain aperture sizes. 
The background flux was estimated by the method described in \S \ref{bg}. 
In creating the background images, several parameters were chosen to give a result that was
representative of the diffuse emission. 

(i) A subgrid size was chosen over which one local background level 
was estimated. The chosen value was 200 pixels, and comparison values are 100, 150, 250, and 300 pixels.
(ii) A clipping factor was chosen so that grids with source area
larger than the clipping factor were omitted. 
The chosen value was 0.3, and comparison values are 0.1, 0.2, 0.4, and 0.5.
(iii) The threshold value for the source flux was chosen. 
All pixels with flux value above the threshold were 
considered source regions. The chosen value was 1.0, and comparison values are 0.5 and 1.5.

\begin{figure}[h]
\center
\includegraphics[scale=0.7]{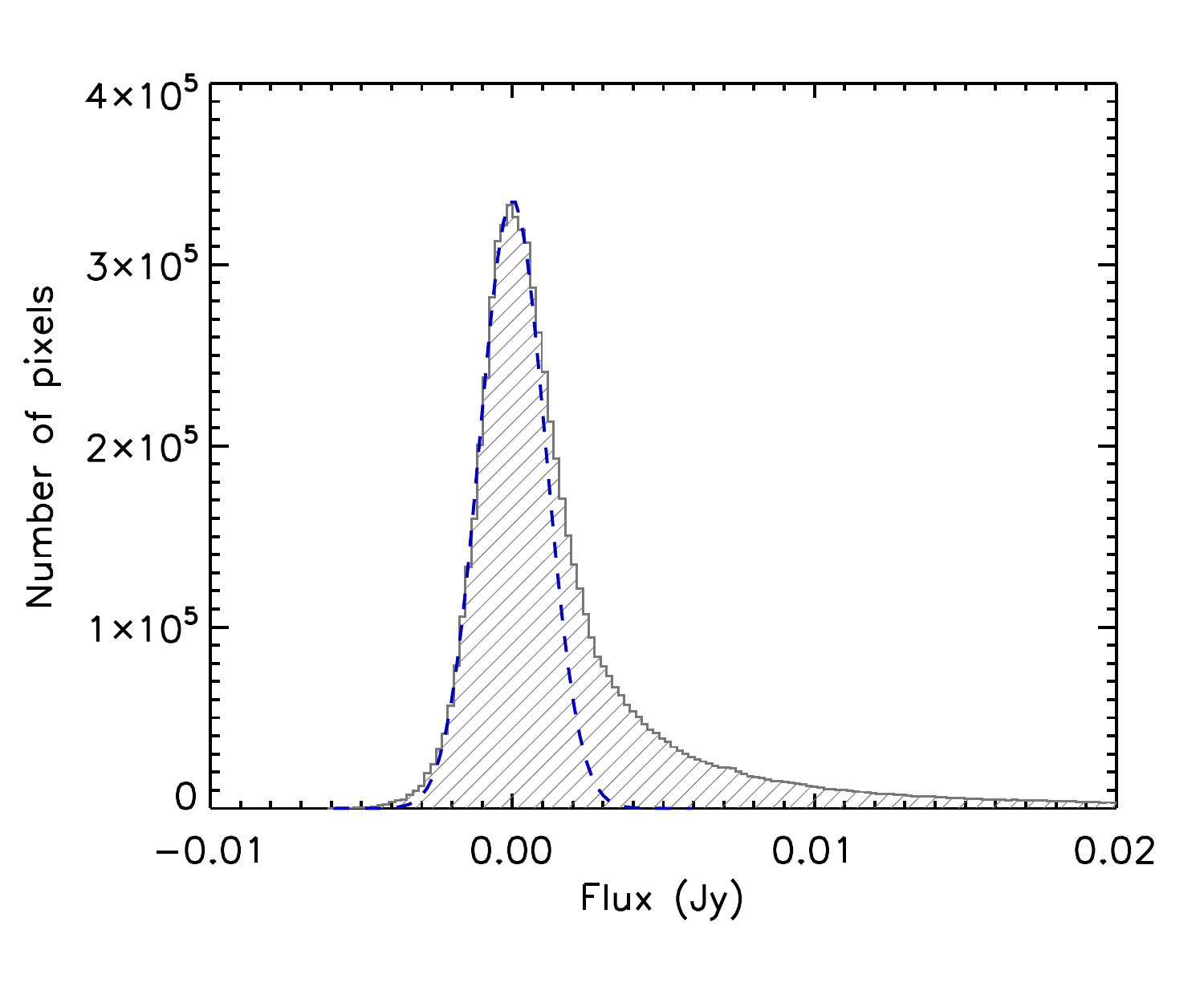}
\caption{Pixel flux distribution for 22 \mic\ image of region 1. The 
blue dashed line shows the Gaussian fit to the flux distribution of flux $<0$ and 
a reflected image of the negative flux for flux $>0$. }
\label{fig:noisefit}
\end{figure}

How the changes in these parameters affect the resulting background 
estimations was considered in the estimations of $\sigma_{\text{B/pix}}$. 
We calculated $\sigma_{\text{B/pix}}$ by changing the three parameters 
around the chosen values and created the background images. We then compared the resulting images
 to the chosen background images.  
The differences between two background images were quantified by
\begin{equation}
\sigma_{\text{diff}}^2 = \frac{1}{N} \sum\limits_{i=1}^N (f_i - f_i')^2, \nonumber
\end{equation}
where N is the number of pixels in the image, $f_i$ is the pixel flux of the chosen image, and $f_i'$ is 
the pixel flux for the comparison image. $\sigma_{\text{diff}}^2 $ were calculated for all the comparing 
background images and the average of the results was adopted as the value for  $\sigma_{\text{B/pix}}^2$.
To see how our estimation of  $\sigma_{\text{B/pix}}$ compared to the spatial variations of the flux, 
we looked at the pixel flux distribution of the background subtracted 22 \mic\ image of region 1.  
Figure~\ref{fig:noisefit} shows the pixel flux distribution of region 1. 
To fit the pixel noise variations, we reflected the negative flux distribution about zero and fitted a Gaussian 
distribution. The resulting fit is shown as the orange dashed line in the figure. The fitting gave a Gaussian 
distribution parameters of $\sigma = 0.001$ Jy/pixel. At the aperture size of 80\arcsec, the noise 
corresponds to $\sigma \approx 0.02$ Jy/source. Our estimation of $\sigma_{\text{B/pix}}$ for region 1
gave $\sigma \approx 0.11$ Jy/source. 

\section{Comparing methods of diffuse IR background estimations}
\begin{figure}[h]
\center
\includegraphics[scale=0.9]{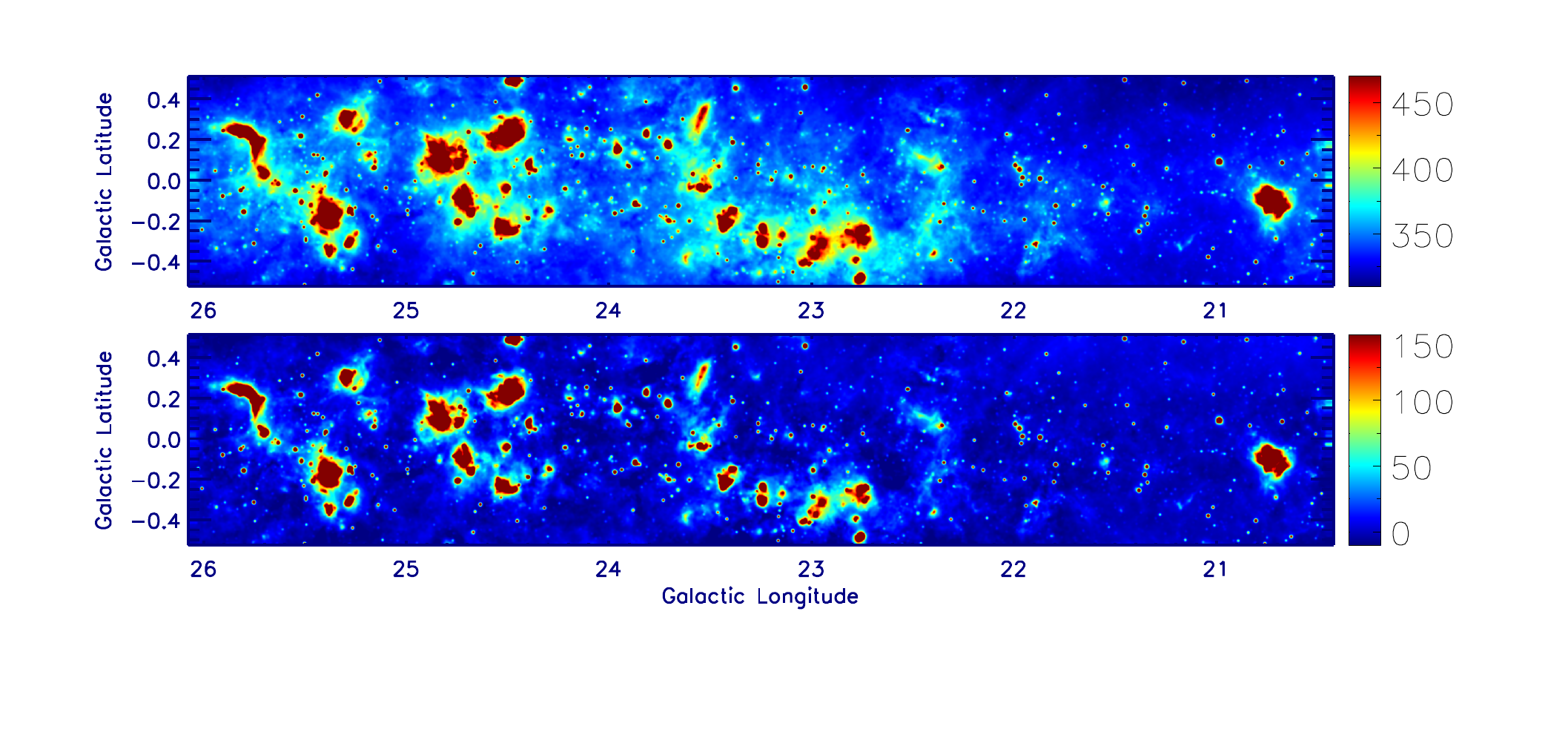}
\caption{WISE 22 \mic\ image of region 1.
The top image shows the original mosaic, and the 
bottom image shows the background-subtracted mosaic.
The color bar is in the unit of MJy/sr. }
\label{fig:bg}
\end{figure}

Our method of estimating diffuse 22 \mic\ emission is described in \S \ref{bg}. 
Figure~\ref{fig:bg} shows the WISE 22 \mic\ images for region 1 before and after 
diffuse background subtraction. 
We compared our method to the cirrus removal method from Battersby et al. (2011, hereafter B11).
B11 estimated diffuse emission for the Herschel 500 \mic\ emission.
The brief summary of the method is as follow. The original image was convolved with 
a Gaussian beam; then a Gaussian fit was performed in the Galactic latitude at each of the 
Galactic longitude. The fitted image was subtracted from the original image. 
The subtracted image was used for estimating a cutoff (4.25$\sigma$) of source flux so that 
everything above the cutoff was considered as sources. Area above the cutoff was masked out 
in the original image.
The process was then iterated until the source mask cutoff converges. 
The original image was masked out with the final cutoff value and convolved with a Gaussian kernel 
to create a background subtracted image. 

\begin{figure}[h]
\center
\includegraphics[scale=0.7]{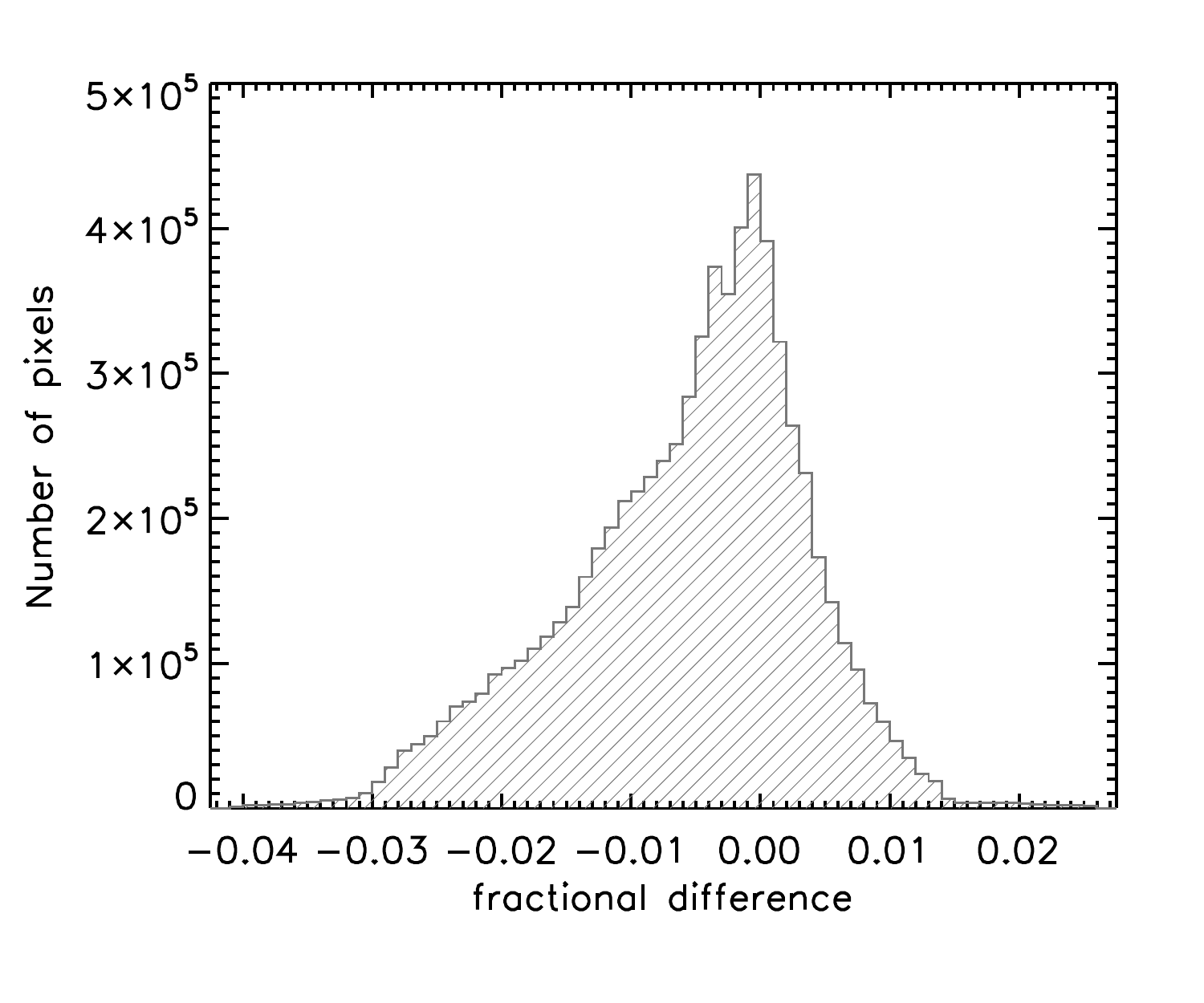}
\caption{A comparison between two methods of background estimations for 
the 22 \mic\ image of region 1. The fractional difference is the ratio of the difference between 
our background image and B11 background image over the original 22 \mic\ image. }
\label{fig:bg_com}
\end{figure}

We compared the background images of region 1 from our method with the method of B11 
performed on the same region of the WISE 22 \mic\ image using a Gaussian kernel FWHM = 12\arcmin.
The result shows that the background image from B11 method gave a comparable background to 
our method with slightly stronger background in bright source area. 
Figure~\ref{fig:bg_com} shows a histogram of (our background image - B11 background image)/(original 22 \mic\ 
image). The fractional differences are small with the highest absolute value of 0.04. 
This result indicates that the choice of a method of background subtraction does not significantly affect the photometric flux. 

\end{document}